\documentclass{IEEEoj}
\usepackage{cite}
\usepackage{amsmath,amssymb,amsfonts}
\usepackage{graphicx,color}
\usepackage{textcomp}
\usepackage{graphicx} 
\usepackage{float}
\usepackage[outdir=./]{epstopdf}
\usepackage{afterpage}
\usepackage{optidef}
\usepackage{orcidlink}
\usepackage{soul}
\usepackage{tikz}
\usetikzlibrary{shapes.geometric, fit, positioning, decorations.pathreplacing}
\usetikzlibrary{trees}
\usepackage{multirow}
\usepackage{subcaption}
\usepackage{algorithm}
\usepackage{algpseudocode}
\usepackage{pgfplots}
\usepackage{comment}
\usepackage{colortbl}

\usepackage{bm} 

\def\BibTeX{{\rm B\kern-.05em{\sc i\kern-.025em b}\kern-.08em
    T\kern-.1667em\lower.7ex\hbox{E}\kern-.125emX}}
\AtBeginDocument{\definecolor{ojcolor}{cmyk}{0.93,0.59,0.15,0.02}}
\def\OJlogo{\vspace{-4pt}\hskip-4pt\includegraphics[height=18pt]{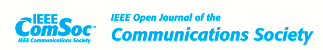}}

\makeatletter
\ams@newcommand{\iiiiint}{\DOTSI\protect\MultiIntegral{5}}
\renewcommand{\MultiIntegral}[1]{%
  \edef\ints@c{\noexpand\intop
    \ifnum#1=\z@\noexpand\intdots@\else\noexpand\intkern@\fi
    \ifnum#1>\tw@\noexpand\intop\noexpand\intkern@\fi
    \ifnum#1>\thr@@\noexpand\intop\noexpand\intkern@\fi
    \ifnum#1>4 \noexpand\intop\noexpand\intkern@\fi 
    \noexpand\intop
    \noexpand\ilimits@
  }%
  \futurelet\@let@token\ints@a
}
\makeatother

\begin{document}
\bstctlcite{IEEEexample:BSTcontrol}
\receiveddate{XX Month, XXXX}
\reviseddate{XX Month, XXXX}
\accepteddate{XX Month, XXXX}
\publisheddate{XX Month, XXXX}

\title{Semantic Communication for Cooperative Multi-Tasking over Rate-Limited Wireless Channels with Implicit Optimal Prior}

\author{Ahmad Halimi Razlighi\,\orcidlink{0009-0006-3826-832X}\IEEEmembership{(Graduate Student Member, IEEE)}, Carsten Bockelmann\,\orcidlink{0000-0002-8501-7324}\IEEEmembership{(Member, IEEE)}, AND Armin Dekorsy\,\orcidlink{0000-0002-5790-1470}
\IEEEmembership{(Senior Member, IEEE)}}

\affil{Department of Communications Engineering, University of Bremen, Germany}
\corresp{CORRESPONDING AUTHOR: Ahmad Halimi Razlighi  (e-mail: halimi@ant.uni-bremen.de).}
\authornote{This work was supported by the German Federal Ministry of Research, Technology, and Space (BMFTR) under Grant 16KISK016 (Open6GHub).}
\markboth{Semantic Communication for Cooperative Multi-Tasking over Rate-Limited Wireless Channels with Implicit Optimal Prior}{Halimi Razlighi \textit{et al.}}

\begin{abstract}
In this work, we expand the cooperative multi-task semantic communication framework (CMT-SemCom) introduced in \cite{10654356}, which divides the semantic encoder on the transmitter side into a common unit (CU) and multiple specific units (SUs), to a more applicable design. Our proposed system model addresses real-world constraints by introducing a general design that operates over rate-limited wireless channels. Further, we aim to tackle the rate-limit constraint, represented through the Kullback-Leibler (KL) divergence, by employing the density ratio trick alongside the implicit optimal prior method (IoPm). By applying the IoPm to our multi-task processing framework, we propose a hybrid-learning approach that combines deep neural networks with kernelized-parametric machine learning methods, enabling a robust solution for the CMT-SemCom. Our framework is grounded in information-theoretic principles and employs variational approximations to bridge theoretical foundations with practical implementations. Simulation results demonstrate the proposed system’s effectiveness in rate-constrained multi-task SemCom scenarios, highlighting its potential for enabling intelligence in next-generation wireless networks.
\end{abstract}

\begin{IEEEkeywords}
Cooperative multi-tasking, deep learning, hybrid learning, information theory, implicit optimal prior, parametric methods, semantic communication.
\end{IEEEkeywords}

\maketitle

\section{Introduction} \label{section.Intro}

Recent advancements in artificial intelligence, particularly in deep learning (DL) and end-to-end (E2E) communication technologies, have led to the rise of \emph{semantic communication} (SemCom) \cite{Gunduz2022,Luo2022,CALVANESESTRINATI2021107930,qin2021semantic}. It has attracted significant attention, being recognized as a critical enabler for the sixth generation (6G) of wireless communication networks. SemCom is expected to play a key role in supporting a wide range of innovative applications that will define 6G connectivity and beyond \cite{WenTong}. This is because emerging applications often have to prioritize task execution over the precise reconstruction of transmitted information at the receiver.

In contrast to conventional communication systems, which are grounded in Shannon's information theory \cite{Shannon1948} and focus on the accurate transmission of symbols, SemCom prioritizes understanding the meaning and goals behind transmitted information. Therefore, designing appropriate communication systems requires moving beyond the traditional focus on precise bit transmission and rethinking the aspects that address communication problems. According to Shannon and Weaver's work, the communication problem is categorized into three levels, each addressing a specific issue \cite{Sana2022}:
\begin{itemize}
    \item The technical problem: Accurate transmission of symbols,
    \item The semantic problem: Transmitting the desired meaning precisely through symbols,
    \item The effectiveness problem: Effectiveness of the received meaning.
\end{itemize}

To meet the demands of emerging applications, SemCom operates at the second level of communication where the goal is to convey the desired meaning rather than ensuring exact bit-level accuracy. By surpassing the traditional focus on the precise transmission of bits, SemCom is well-suited for new applications, such as the industrial internet and autonomous systems, where successful task execution is prioritized over the exact reconstruction of transmitted data at the receiver.

Research into SemCom has explored five main approaches, with four detailed in \cite{Wheeler2023} and a fifth inspired by Weaver's extension of Shannon’s theory to include the semantic level \cite{weaver1953recent}. These approaches are:
\begin{itemize}
    \item  Classical approach,
    \item Knowledge graph approach,
    \item Machine learning (ML) approach,
    \item Significance approach,
    \item Information theory approach.
\end{itemize} 

The classical approach utilizes \emph{logical} probability to quantify semantic information. Bar-Hillel and Carnap \cite{BarHillelCarnap}, introduced this approach and have inspired many other works introducing methods to measure the semantic information of a source. As noted in \cite{Wheeler2023}, this definition of semantic information primarily applies to psychological investigations rather than communication counterparts. 

Next, the knowledge graph approach represents semantics by knowledge graph structures. This approach stores the information such that the semantic relations between entities are held via semantic matching models as a knowledge graph technique \cite{Wang2017}. For instance, \cite{Zhou2022} exploits this approach for its proposed semantic information detection framework, using triplets of the graph as semantic symbols. 

The ML approach leverages learned model parameters to represent semantics. The ML approach lacks the communication-theoretic analysis in the semantic communication domain, relying on defined loss functions and the black-box nature of its tools, such as deep neural networks (DNNs).
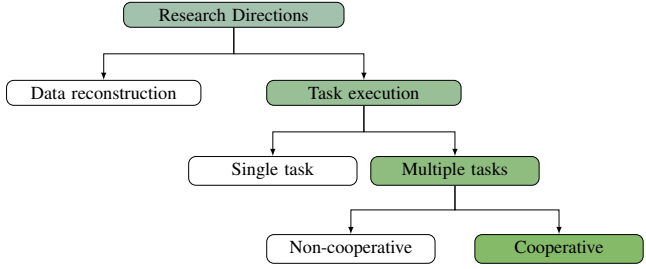
\begin{figure}[!t]
    \centering
    \resizebox{0.5\textwidth}{!}{\definecolor{cambridgeblue}{rgb}{0.64, 0.76, 0.68}
\definecolor{cambridgeblue2}{rgb}{0.6, 0.75, 0.59}
\definecolor{cambridgeblue3}{rgb}{0.56, 0.74, 0.50}
\definecolor{dollarbill}{rgb}{0.52, 0.73, 0.41}

\begin{tikzpicture}[
  edge from parent fork down,
  every node/.style = {rounded corners, draw, align=center, text width=3cm},
  level 1/.style = {sibling distance=5cm, level distance=1.5cm},
  level 2/.style = {sibling distance=3.5cm, level distance=1.5cm},
  level 3/.style = {sibling distance=4cm, level distance=1.5cm},
  edge from parent/.style={draw=black, -latex}
  ]

  \node[fill=cambridgeblue,text width=4cm] {Research Directions} 
    child {node[text width=3.5cm] {Data reconstruction}}
    child {node[fill=cambridgeblue2,text width=3.5cm] {Task execution}
      child {node[draw=black,text width=3cm] {Single task}}
      child {node[fill=cambridgeblue3,draw=black,text width=3cm] {Multiple tasks}
        child {node[draw=black,text width=3cm] {Non-cooperative}}
        child {node[fill=dollarbill,draw=black,text width=3cm] {Cooperative}}
      }
    };

\end{tikzpicture}}
    \caption{Categorization of research works in semantic communications.}
    \label{fig:research_directions}
\end{figure}

The significance approach considers the significance of information as its semantics. Although it is argued in \cite{Wheeler2023} that this approach is more about investigating the effectiveness problem of communication, its application for the semantic problem has been studied emphasizing \emph{timing} as semantics. This is specifically explored in the semantic communication domain under the Age of Information (AoI) topic \cite{Gunduz2022}. 

Lastly, inspired by Weaver, an alternative approach extends Shannon's \emph{statistical} probability (information theory) beyond the technical layer to the next two levels. Recently, some works have adopted information theory in investigating semantic communication, which are mentioned later.

The research work in SemCom primarily focuses on two research directions: data reconstruction and task execution, illustrated in Fig. \ref{fig:research_directions}. ‌‌‌‌Initial investigations into data recovery were led by \cite{Xie2021} and \cite{Xie2021-2}, which utilized the ML approach to reconstruct diverse data sources such as text, speech, and images. Building on these foundational works, \cite{10599525MBM,xu2024semanticawarepowerallocationgenerative,Yan2022,Wang2022,Tong2021} have extended the focus to explore communication concepts like efficiency and resource allocation in SemCom. In addition, SemCom systems dealing with structured data have been examined through the knowledge graph approach to enhance data recovery \cite{9685056KG}. Recent developments also address robustness and reliability, for instance, \cite{JammingSemantic} proposed a multi-functional reconfigurable intelligent surface-assisted framework for semantic anti-jamming communication, showing how semantic transceivers and RIS can jointly provide resilience.

On the other hand, task-oriented communication or goal-oriented communication can be categorized into single-task processing and multi-task processing. The latter is further divided into two directions: non-cooperative processing and cooperative multi-task processing. Our paper specifically addresses cooperative multi-task processing within the context of SemCom. A review of the literature related to task execution SemCom is provided in Subsection \ref{subsec:related_works}.

To better situate our contribution, we note that there exist generally two main paradigms in multi-task learning \cite{wu2020understandingMTL}: (i) multiple tasks on different datasets, and (ii) multi-label tasks where different classification tasks are supported based on a single dataset. Our focus falls into the second category, focusing on a multi-label domain, where multiple semantic variables are extracted from the same observation and represent different classification tasks. This perspective is elaborated in detail in Section \ref{section.SystemModel}.

\subsection{Related Works} \label{subsec:related_works}

In task-oriented SemCom, the focus shifts to executing intelligent tasks at the receivers. Most research in this area has concentrated on single-task scenarios. For example, \cite{Shao2021} developed a communication scheme using the information bottleneck (IB) framework, which encodes information while adapting to dynamic channel conditions. Moreover, the same authors in \cite{Shao2022} studied distributed relevant information encoding for collaborative feature extraction to fulfill a single task. \cite{Beck2023} also offered a framework for collaborative retrieval of the message using multiple received semantic information. Recent studies have highlighted the integration of communication with computation and sensing (ISCC) in this context. In particular, \cite{Task_SensingComm} investigated a multi-device edge inference with ISCC  for improved inference accuracy in a classification task.

To address practical communication scenarios, SemCom systems must be capable of handling multiple tasks simultaneously. Early efforts, such as \cite{xie2022task,he2022learning}, explored non-cooperative methods where each task operates on its respective dataset independently. Conversely, recent works like \cite{10013075,10520522,gong2023scalable} studied joint multi-tasking using established ML approaches and architecture \cite{caruana1997multitask} for SemCom systems. Although these works incorporated communication aspects like channel conditions in their studies, their multi-task processing is based exclusively on ML approaches. 

On the other hand, in \cite{10654356}, we introduced an information-theoretic analysis of a cooperative multi-task (CMT) SemCom system, avoiding the black-box use of DNN. \cite{10654356} investigated a split structure for the semantic encoder, dividing the semantic encoder into a common unit (CU) and multiple specific units (SUs), to enable cooperative processing of various tasks on the transmitter side. The proposed CMT-SemCom can perform multi-tasking based on a single observation. Further, \cite{halimirazlighi2024} expanded the CMT-SemCom to scenarios, in which, instead of full observation, distributed partial observations are available. By introducing CCMT-SemCom for multi-tasking in \cite{halimirazlighi2024}, we combined the cooperative processing on the transmitter side with the collaborative processing, where multiple nodes collaborate to execute their shared task, on the receive side. In addition, on exploring the physical layer communications aspects, \cite{10502352TWC} has studied resource allocation for multi-task SemCom networks.  
    
\subsection{Motivations and Contributions}
This work builds upon the CMT-SemCom framework introduced in \cite{10654356} and extends it to a more realistic setting by incorporating rate-limited wireless communication channels. The presence of this constraint introduces a Kullback-Leibler (KL) divergence term in the objective function of the specific units, which must be handled during the learning step.

To better address this constraint, we propose a separation-based design where the CU and the SUs are optimized in turn. This not only clarifies their distinct functional roles but also leads to a more tractable formulation of the constrained learning problem. In addition, as shown in recent research works, i.e., \cite{10599525MBM,halimirazlighi2024}, such a separation-based design offers better compatibility with handling different channel conditions by reducing the number of trained parameters for each channel condition.

Existing approaches rely on a fixed prior when regularizing the KL term, e.g., \cite{kingma2013auto,Shao2021}, which can limit the flexibility and performance of the system. In contrast, we propose to adopt the Implicit Optimal Prior method (IoPm) in this work, which leverages density ratio estimation to better approximate the prior in a data-driven manner. However, while investigating, we found that directly integrating IoPm into a fully DNN-based implementation of CMT-SemCom proves ineffective due to the challenges of instability.

To overcome this, we introduce a hybrid-learning strategy that combines deep neural networks with kernelized-parametric machine learning techniques. This allows us to effectively implement IoPm while preserving the benefits of our cooperative multi-task semantic communication framework.

In summary, key contributions are:

\begin{itemize}
    \item Extending the CMT-SemCom system to operate under rate-limited wireless channels, reflecting practical communication constraints.
    \item Proposing a separation-based design of the CU and SUs to achieve a more structured and effective formulation for constrained optimization.
    \item Addressing the limitations of fixed-prior regularization by adopting IoPm for more flexible and accurate KL divergence approximation.
    \item Introducing a hybrid-learning approach that integrates DNNs with parametric ML to robustly implement IoPm within the CMT-SemCom framework.
\end{itemize}
\begin{table}[!t]
    \centering
    \caption{The Table of Notations.}
    \begin{tabular}{c c}
        \hline
        \textbf{Notation} & \textbf{Definition} \\
        \hline
        $\mathbf{S}$ & observation (input) \\
        $\mathbf{z}$ & semantic variables \\
        $\mathbf{c}$ & output of the CU \\
        $\mathbf{x}_n$ & output of the n-th SU encoder \\
        $\mathbf{n}$ & additive white Gaussian noise\\
        $\mathbf{\hat{x}}_n$ & noise-corrupted version of  $\mathbf{x}_n$  \\
        $I(\cdot ; \cdot)$ & mutual information \\
        $KL(\cdot \parallel \cdot)$ & Kullback-Leibler divergence \\
        $\mathbb{E}[\,\cdot\,]\,$ & expectation \\
        $\mathcal{L}(\cdot)$ & objective function \\
        $\boldsymbol{\theta}$ & neural network (NN) parameters of the CU encoder \\
        $\boldsymbol{\Xi}$ & NN parameters of the auxiliary CU decoders\\
        $\boldsymbol{\phi}_n$ & NN parameters of the n-th SU encoder\\
        $\boldsymbol{\psi}_n$ & NN parameters of the n-th SU decoder\\
        $\boldsymbol{\mu}$, $\boldsymbol{\sigma}$ & mean and standard deviation of a Gaussian distribution\\
        $\mathbf{\epsilon}$ & auxiliary random variable for reparameterization trick \\
        $r(\cdot)$ & density ratio function \\
        $\boldsymbol{\omega}$ & parameter vector of the density ratio estimator \\
        $\Omega(\cdot)$ & basis function \\
        $K(\cdot , \cdot)$ & kernel function \\
        $\sigma_k$ & kernel bandwidth \\
        \hline
    \end{tabular}
    \label{table_notation}
\end{table}

\subsection{Organization and Notations}

The rest of the paper is organized as follows. Section \ref{section.SystemModel} presents probabilistic modeling of the proposed system model, followed by presenting two distinct objective functions that enable the separation-based design of the CU and SUs in \ref{subsec.CU_loss_function} and \ref{subsec.SU_loss_function}, respectively. Next, \ref{sub:IoP} describes the IoP method for enhanced approximation of the constrained problem in the SUs objective function and the proposed hybrid-learning approach. Section \ref{section.Simulation} presents simulation results evaluating the performance of the proposed CMT-SemCom across various datasets. Finally, section \ref{section.Conclusion} concludes the paper highlighting the key findings. We also note that the notations used throughout this paper are listed in Table \ref{table_notation}.

\section{System Model} \label{section.SystemModel}

This section explores the separation-based design for the proposed CMT-SemCom system model under constrained wireless channels. We begin by presenting the probabilistic modeling of the proposed framework in \ref{subsec.system_prob_model}. Following this, we formulate two distinct optimization problems: one focusing on the design of the CU, responsible for promoting cooperation amongst tasks, and the other targeting the design of the SUs, which is responsible for the joint semantic and channel coding (JSCC). We adopt the information maximization (Infomax) principle in \ref{subsec.CU_loss_function}, while employing the information bottleneck (IB) approach in \ref{subsec.SU_loss_function} to formulate the objective function for our constrained optimization problem. Next, \ref{sub:IoP} presents the IoPm and our hybrid-learning approach. 

\subsection{System Probabilistic Modeling}\label{subsec.system_prob_model}

\begin{figure}[!t]
    \centering
    \resizebox{0.5\textwidth}{!}{\begin{tikzpicture}[>=stealth, node distance=1.5cm]
        \node[text=red] (G) {};
        \node[right=3 of G, text=red] (Z) {};
        \node[right=0.32 of G] (B) {};
        \node[left=0.3 of Z] (Q) {};
        \node[right=2.5 of Z, yshift=0.3 cm] (S) {$\begin{tabular}{@{}c@{}}
                                                    \text{Observation} \\
                                                    \small \text{($\bm{S}$)}
                                                  \end{tabular}$};
        \draw[fill=white] (G.center)++(0.8,0.8) circle (0.5) node[font=\scriptsize] (GCn) {$\text{Task}_N$};
        \draw[fill=white] (Z.center)++(0.8,0.8) circle (0.5) node[font=\small] (ZCn) {$z_N$};
        \node[right=0.13 of G, xshift=0.8 cm, yshift=0.8 cm] (Bn) {};
        \node[left=0.1 of Z, xshift=0.8 cm, yshift=0.8 cm] (Qn) {};
        \draw[->] (Bn) -- (Qn);
        \draw[fill=white] (G.center)++(0.3,0.3) circle (0.6) node (GC3) {};
        \draw[fill=white] (Z.center)++(0.3,0.3) circle (0.6) node (ZC3) {};
        \node[right=0.23 of G, xshift=0.3 cm, yshift=0.3 cm] (B2) {};
        \node[left=0.3 of Z, xshift=0.3 cm, yshift=0.3 cm] (Q2) {};
        \draw[->] (B2) -- (Q2);
        \draw[fill=white] (G.center)++(0.15,0.15) circle (0.65) node (GC2) {};
        \draw[fill=white] (Z.center)++(0.15,0.15) circle (0.65) node (ZC2) {};
        \node[right=0.28 of G, xshift=0.15 cm, yshift=0.15 cm] (B1) {};
        \node[left=0.3 of Z, xshift=0.15 cm, yshift=0.15 cm] (Q1) {};
        \draw[->] (B1) -- (Q1); 
        \draw[fill=white] (G.center) circle (0.7) node[font=\small] (GC1) {$\text{Task 1}$};
        \draw[fill=white] (Z.center) circle (0.7) node[font=\large] (ZC1) {$z_1$};
        \draw[->] (B) -- (Q);

        \fill (G.center)++(2.0,0.5) circle [radius=0.7pt];
        \fill (G.center)++(2.07,0.57) circle [radius=0.7pt];
        \fill (G.center)++(2.14,0.64) circle [radius=0.7pt];

        \fill (G.center)++(5.05,0.5) circle [radius=0.7pt];
        \fill (G.center)++(5.12,0.57) circle [radius=0.7pt];
        \fill (G.center)++(5.19,0.64) circle [radius=0.7pt];

        \node[right=0.32 of Z] (QQ) {};
        \node[right=2.47 cm of Z] (BB) {};
        \draw[->] (QQ) -- (BB);
        \node[right=0.29 of Z, xshift=0.15 cm, yshift=0.15 cm] (QQ1) {};
        \node[right=2.33 of Z, xshift=0.15 cm, yshift=0.15 cm] (BB1) {};
        \draw[->] (QQ1) -- (BB1);
        \node[right=0.24 of Z, xshift=0.3 cm, yshift=0.3 cm] (QQ2) {};
        \node[right=2.18 of Z, xshift=0.3 cm, yshift=0.3 cm] (BB2) {};
        \draw[->] (QQ2) -- (BB2);
        \node[right=0.13 of Z, xshift=0.8 cm, yshift=0.8 cm] (QQn) {};
        \node[right=1.68 of Z, xshift=0.8 cm, yshift=0.8 cm] (BBn) {};
        \draw[->] (QQn) -- (BBn);
        \node[draw, rectangle, fit=(S), inner sep=1pt, minimum height=2 cm] (RectangleS) {};
        \node[left=0.35 of Z] (ZSS) {};
        \node[draw, dashed, fit=(RectangleS)(ZSS), inner sep=5pt] (SS) {};
        \node[above=0.1cm of SS, font=\small] {Semantic Source};
\end{tikzpicture}}
    \caption{Probabilistic graphical modeling of the semantic source.}
    \label{fig:semantic_source_model}
\end{figure}
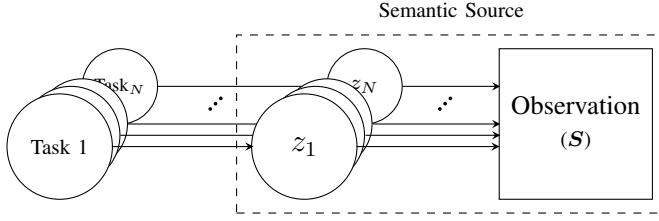

We begin by presenting our interpretation of the \emph{semantic source} concept as discussed in \cite{10654356}. We assume the existence of $N$ independent tasks. Each task is entailed with its specific \emph{semantic variable}, thus we have $N$ semantic variables indicated by $\bm{z}=[\,z_1\, z_2\,\dots\,z_N]\,$. We assume that our semantic variables are entailed with an observation, $\bm{S}$. We define the tuple of $(\bm{z}, \bm{S})$ as our semantic source, fully described by the probability distribution of $p(\bm{z}, \bm{S})$. Fig.~\ref{fig:semantic_source_model} illustrates our interpretation using probabilistic graphical modeling \cite{koller2009probabilistic} and a stack view for a better illustration. Such a definition enables the simultaneous extraction of multiple semantic variables based on a single observation. For instance, consider an image featuring both a tree and a number. One task may entail determining the presence of a tree, resulting in a binary semantic variable. Meanwhile, another task could focus on identifying the number within the image, yielding a multinomial semantic variable.

In this paper, we assume $N$ tasks specify semantic variables to be delivered to their respective recipients through semantic decoders leveraging task-relevant information extracted by CU and SUs. It was demonstrated that when semantic variables share statistical relationships, CMT-SemCom enables cooperative processing and significantly improves performance in multi-task cases by utilizing common information \cite{10654356}. 

Our system model has, on the transmitter side, the encoder split into one CU and multiple SUs. The CU encoder outputs a representation $\bm{c}$, which is the common relevant information extracted from the semantic source, via $p^{\text{\tiny CU}}(\bm{c}|\bm{S})$. Next, each SU$_n$ encodes $\bm{c}$ into a task-specific information $\bm{x}_n$ using $p^{\text{\tiny SU}_n}(\bm{x}_n|\bm{c})$. These channel inputs are then transmitted through a rate-limited additive white Gaussian noise (AWGN) channel, resulting in received signals $\bm{\hat{x}}_n = \bm{x}_n + \bm{n}$, where $\bm{n} \sim \mathcal{N}(\bm{0}_d, \sigma^2_n \bm{I}_d)$, and $d$ indicates the number of channel uses (the limiting constraint) or ‌in other words, the size of the encoded task-specific information, $\bm{x}_n \in \mathbb{R}^{d_n\times1}$. On the Rx side, the semantic decoder $p^{\text{\tiny Dec}_n}(\hat{z}_n|\bm{\hat{x}}_n)$ delivers the semantic variable $z_i$ from $\bm{\hat{x}}_n$. The system model is also illustrated in Fig. \ref{fig:system-model}.

Subsequently, the Markov representation of our system model for the $n$-th semantic variable is outlined as follows:
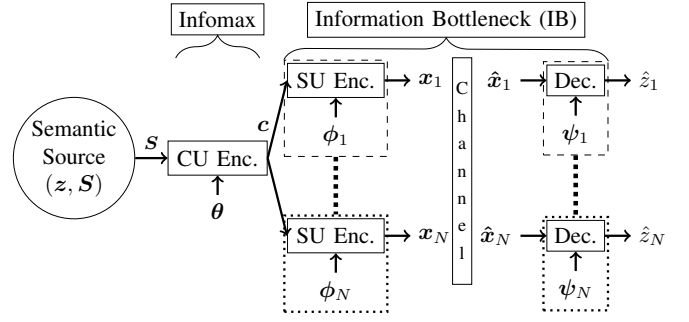
\begin{figure}[!t]
    \centering
    \resizebox{0.5\textwidth}{!}{\begin{tikzpicture}
        \node[draw, circle, inner sep=0.5pt] (S) {$\begin{tabular}{@{}c@{}}
                                                    \text{Semantic} \\
                                                    \text{Source} \\
                                                    \text{$(\bm{z}, \bm{S})$}
                                                  \end{tabular}$};
        \node[draw, rectangle, right=0.5 cm of S, inner sep=4pt] (C) {$\text{CU Enc.}$};
        \node[below=0.3cm of C] (th) {$\boldsymbol{\theta}$};
        \node[draw, rectangle, right=0.3 cm of C, yshift=1.2 cm, inner sep=4pt] (SU1) {$\text{SU Enc.}$};
        \node[below=0.3cm of SU1] (ph1) {$\boldsymbol{\phi}_1$};
        \node[draw, rectangle, dashed, fit=(SU1)(ph1), inner sep=1.2pt](outerSU1){};
        \node[draw, rectangle, right=0.3 cm of C, yshift=-1.2 cm, inner sep=4pt] (SU2) {$\text{SU Enc.}$};
        \node[below=0.3cm of SU2] (ph2) {$\boldsymbol{\phi}_N$};
        \node[draw, dotted, line width=1pt, fit=(SU2)(ph2), inner sep=1.2pt](outerSU2){};
        \node[right=0.4 cm of SU1](X1){$\bm{x}_1$};
        \node[right=0.4 cm of SU2](Xn){$\bm{x}_N$};
        \node[right=of SU2, yshift=1 cm, xshift=-0.02 cm] (ch) {$\begin{tabular}{@{}c@{}}
              \footnotesize C\\
              \footnotesize h\\
              \footnotesize a\\
              \footnotesize n\\
              \footnotesize n\\
              \footnotesize e\\
              \footnotesize l
        \end{tabular}$};
        \node[draw, fit=(ch), minimum height=3.5 cm, inner sep=-2pt]{};
        
        \node[draw, rectangle, right=2.5 of SU1] (Rx1) {$\text{Dec.}$};
        \node[below=0.3cm of Rx1] (ps1) {$\boldsymbol{\psi}_1$};
        \node[draw, dashed, fit=(Rx1)(ps1), inner sep=1.3pt] (RxSU1) {};
        \node[draw, rectangle, right=2.5 of SU2] (Rx2) {$\text{Dec.}$};
        \node[below=0.3cm of Rx2] (ps2) {$\boldsymbol{\psi}_N$};
        \node[draw, dotted, line width=1pt, fit=(Rx2)(ps2), inner sep=1.3pt] (RxSU2) {};
        \node[left=0.4 cm of Rx1](X1_hat){$\bm{\hat{x}}_1$};
        \node[left=0.4 cm of Rx2](Xn_hat){$\bm{\hat{x}}_N$};
        \node[right=0.4 cm of Rx1](Z1_hat){$\hat{z}_1$};
        \node[right=0.4 cm of Rx2](Zn_hat){$\hat{z}_N$};

        \node[above=0.03 cm of C, xshift= 0.7 cm]{$\bm{c}$};

        \draw [decorate, decoration={brace, amplitude=10pt}] (outerSU1.north west) -- (RxSU1.north east) node [black, midway, above=10pt, draw, rectangle] {Information Bottleneck (IB)};
        \node[above=1.15 cm of C.north east] (modal_brace1){};
        \node[above=1.15 cm of C.north west] (modal_brace2){};
        \draw [decorate, decoration={brace, amplitude=10pt, mirror}] (modal_brace1) -- (modal_brace2) node [black, midway, above=10pt, draw, rectangle] {Infomax};
        
        \draw[dotted, line width=2pt] (outerSU1) -- (outerSU2);
        \draw[dotted, line width=2pt] (RxSU1) -- (RxSU2);
        \draw[->, line width=1pt] (S) -- node[above]{\small$\bm{S}$} (C.west);
        \draw[->, line width=1pt] (C.east) -- (SU1.west);
        \draw[->, line width=1pt] (C.east) -- (SU2.west);
        \draw[->, line width=1pt] (SU1) -- (X1);
        \draw[->, line width=1pt] (SU2) -- (Xn);
        \draw[->, line width=1pt] (X1_hat) -- (Rx1);
        \draw[->, line width=1pt] (Xn_hat) -- (Rx2);
        \draw[->, line width=1pt] (Rx1) -- (Z1_hat);
        \draw[->, line width=1pt] (Rx2) -- (Zn_hat);
        \draw[->, line width=1pt] (th) -- (C);
        \draw[->, line width=1pt] (ph1) -- (SU1);
        \draw[->, line width=1pt] (ph2) -- (SU2);
        \draw[->, line width=1pt] (ps1) -- (Rx1);
        \draw[->, line width=1pt] (ps2) -- (Rx2);
        
\end{tikzpicture}}
    \caption{Illustration of the proposed separation-based design for the CMT-SemCom framework under rate-limit wireless channels.}
    \label{fig:system-model}
\end{figure}
\begin{equation}
\begin{split}
    &p(\hat{z}_n,\bm{\hat{x}}_n,\bm{x}_n,\bm{c}|\bm{S}) =\\[0.3em] & p^{\text{\tiny Dec$_n$}}(\hat{z}_n|\bm{\hat{x}}_n)\,p^{\text{\tiny Channel}}(\bm{\hat{x}}_n|\bm{x}_n)\,p^{\text{\tiny SU$_n$}}(\bm{x}_n|\bm{c})p^{\text{\tiny CU}}(\bm{c}|\bm{S}).
\end{split}
\label{eq:system_probability}
\end{equation}

\subsection{CU Objective Function} \label{subsec.CU_loss_function}

To begin the separation-based design for the CU, which is shared amongst all SUs, we formulate the following optimization problem, adopting the Infomax principle.
\begin{equation} \label{eq.CU-optimization}
    p^{\text{\tiny CU}}(\bm{c}|\bm{S})^\star = \arg \max_{p^{\text{\tiny CU}}(\bm{c}|\bm{S})} I(\bm{c};\bm{z}) .
\end{equation}

Hence, our objective is to maximize the mutual information between the CU output,$\bm{c}$, and the underlying semantic variables, $\bm{z}=[\,z_1\, z_2\, \dots\, z_N]\,^T$, associated with the observation. Considering the availability of a sample set instead of the true distribution for $p(\bm{S},\bm{z})$, we approximate the semantic source distribution with the corresponding available sample set \cite{bishop2006pattern}. Moreover, we employ the variational method, which is a way to approximate intractable computations based on some adjustable parameters, like weights in neural networks (NNs) \cite{kingma2013auto}. The technique is widely used in machine learning, e.g., \cite{alemi2016deep}, and also in task-oriented communications, e.g., \cite{Shao2021} and \cite{Shao2022}. Thus, we approximate the posterior distribution $p^{\text{\tiny CU}}(\bm{c}|\bm{S})$ by variational approximation using NN parameterized by $\boldsymbol{\theta}$. This approximation yields $p^{\text{\tiny CU}}_{\boldsymbol{\theta}}(\bm{c}|\bm{S})$ and we present the CU objective function as follows.
\begin{equation} \label{eq:CU-objective}
    \begin{aligned}
        \mathcal{L}^{\text{\tiny CU}}(\boldsymbol{\theta}) &\approx I(\bm{c};\bm{z}) \\[0.6em]
        &\approx \int p(\bm{S},\bm{z})\,p^{\text{\tiny CU}}_{\boldsymbol{\theta}}(\bm{c}|\bm{S}) \log p(\bm{z}|\bm{c})\, d\bm{S}\, d\bm{z}\, d\bm{c} \\[0.6em]
        &\approx \mathbb{E}_{p^{\text{\tiny CU}}_{\boldsymbol{\theta}}(\bm{c}|\bm{S})} \left[\ \mathbb{E}_{p(\bm{S},\bm{z})} [\ \log p(\bm{z}|\bm{c})\, ]\,\right]\,.
    \end{aligned}
\end{equation}

The outer expectation shows how the CU integrates the common knowledge extraction amongst the SUs and emphasizes our distinct approach caused by our architecture in cooperative processing. In addition, a detailed derivation for the infomax objective function of the CU can be found in \cite{10654356}. Given the availability of the semantic source, denoted by the joint probability distribution $p(\bm{S},\bm{z})$ and the posterior distribution $p^{\text{\tiny CU}}_{\boldsymbol{\theta}}(\bm{c}|\bm{S})$, the semantic space posterior $p(\bm{z}|\bm{c})$ could be fully determined:
\begin{equation} \label{eq:CU-log-likelihood}
        p(\bm{z}|\bm{c}) = \int\frac{p(\bm{S},\bm{z})\,p^{\text{\tiny CU}}_{\boldsymbol{\theta}}(\bm{c}|\bm{S})\,}{p_{\boldsymbol{\theta}}(\bm{c})}\,d\bm{S}.
\end{equation}
\algrenewcommand\algorithmicrequire{\textbf{Input:}}
\algrenewcommand\algorithmicensure{\textbf{Output:}}
\algrenewcommand\algorithmicdo{}
\begin{algorithm}[!t]
    \caption{Training the CU Encoder.} \label{algorithm:CU_training}
    \begin{algorithmic}[1]
        \Require \text{Preprocessed training dataset:} $\{\bm{S}^{(m)},z^{(m)}_{1},\dots, z^{(m)}_{N}\}_{m\in M}$, number of iterations $T$, batch sizes $M_n$
        \Ensure The trained parameters $\boldsymbol{\theta}$ and $\boldsymbol{\Xi}$
        \For{epoch $t=1$ to $T$}
        \For{$n=1$ to $N$}
            \State \text{Randomly select a minibatch $\{(\bm{S}^{(m)},z_{n}^{(m)}) \}_{m=1}^{M_n}$}
            \State \parbox[t]{\dimexpr\linewidth-\algorithmicindent}{Compute the mean vector $\{\boldsymbol{\mu}_{\bm{c}|\bm{S}^{m}}\}_{m=1}^{M_n}$}
            \State \text{Compute the standard deviation vector $\{\boldsymbol{\sigma}_{\bm{c}|\bm{S}^{m}}\}_{m=1}^{M_n}$}
            \For{$m=1$ to $M_n$}
                \State \text{Sample the $\{\boldsymbol{\epsilon}^{l}\}_{l=1}^{L} \sim \mathcal{N}(\mathbf{0}, \mathbf{I})$}
                \State \text{Compute $\bm{c}^{(m,l)}= \boldsymbol{\mu}_{\bm{c}|\bm{S}^{m}} + \boldsymbol{\sigma}_{\bm{c}|\bm{S}^{m}} \odot \boldsymbol{\epsilon}^l$}
            \EndFor    
            \State \text{Compute the log-likelihood $\log q_{\boldsymbol{\xi}_n}(z_n|\mathbf{c}^{(m,l)})$}
        \EndFor
        \State Compute the loss $\mathcal{L}^{\text{\tiny CU}}$ based on (\ref{eq:CU-objective-empirical}).
        \State Update parameters $\boldsymbol{\theta}$ and $\boldsymbol{\Xi}$ through backpropagation.
        \EndFor
    \end{algorithmic}
\end{algorithm}
Considering that \mbox{$p_{\boldsymbol{\theta}}(\bm{c})=\int p^{\text{\tiny CU}}_{\boldsymbol{\theta}}(\bm{c}|\bm{S})\,p(\bm{S}) \,d\bm{S}$} could be also available. However, due to intractability of high dimensional integrals, we apply another variational approximation, replacing the true posterior distribution with its approximation $p_{\boldsymbol{\Xi}}(\bm{z}|\bm{c})$ where $\boldsymbol{\Xi}=[\,\boldsymbol{\xi}_1\, \boldsymbol{\xi}_2\, \dots\, \boldsymbol{\xi}_N]\,^T$ is the parameters of the corresponding NNs of the auxiliary decoders for training the CU. Thus, the objective function in (\ref{eq:CU-objective}), is expressed as below.
\begin{equation} \label{eq:CU-objective-final}
    \mathcal{L}^{\text{\tiny CU}}(\boldsymbol{\theta}, \boldsymbol{\Xi}) \approx \mathbb{E}_{p^{\text{\tiny CU}}_{\boldsymbol{\theta}}(\bm{c}|\bm{S})} \left[\ \mathbb{E}_{p(\bm{S},\bm{z})} [\ \log p_{\boldsymbol{\Xi}}(\bm{z}|\bm{c})\, ]\,\right]\,.
\end{equation}

Further, we approximate the expectations with Monte Carlo sampling following data-driven approach, given that there exists a dataset $\{\mathbf{S}^{(m)},z^{(m)}_{1},\dots, z^{(m)}_{N}\}^M_{m=1}$ where $M$ represents the dataset size and $N$ denotes the number of available tasks. Thus, the empirical estimation of the objective function can be expressed as:
\begin{equation} \label{eq:CU-objective-empirical}
    \mathcal{L}^{\text{\tiny CU}}(\boldsymbol{\theta}, \boldsymbol{\Xi}) \approx \frac{1}{L}\sum_{l=1}^{L}\left[\,\sum_{n=1}^{N}\bigg\{\frac{1}{M_n}\sum_{m=1}^{M_n} \log p_{\boldsymbol{\xi}_n}(z_n|\bm{c}^{(m,l)})\,\bigg\}\right]\,.
\end{equation}

Additionally, we have applied the reparameterization trick \cite{kingma2022autoencoding}, to overcome the differentiability issues in the backpropagation of the objective function by introducing $\bm{c}^{(m,l)}= \boldsymbol{\mu}_{\bm{c}|\bm{S}^{m}} + \boldsymbol{\sigma}_{\bm{c}|\bm{S}^{m}} \odot \boldsymbol{\epsilon}^l$ where the auxiliary variable $\boldsymbol{\epsilon}^l \sim \mathcal{N}(\mathbf{0},\mathbf{I})$ and $\odot$ represents the element-wise product. Details on how the CU loss function is differentiable with respect to (w.r.t) $\boldsymbol{\theta}$ and the reparameterization trick are deferred to the Appendix \ref{Appendix_CU-reparameterization}. In (\ref{eq:CU-objective-empirical}), $L$ is the sample size of the reparameterization trick, fixed to one, signifying that the CU updates once, encompassing $N$ specific features. $M_n$ appears for the minibatch size of $\{(\mathbf{S}^{(m)},z_{n}^{(m)})\}_{m=1}^{M_n}$ and for simplicity we assume that the minibatch sizes are equal across semantic variables, $M_n = M$. The training procedure for the CU is described in Algorithm~\ref{algorithm:CU_training}.

\subsection{SU Objective Function} \label{subsec.SU_loss_function}

SUs are responsible for the JSCC, transmitting task-specific information such that the respective recipients can decode the intended semantic variables. To design the SU concerning the rate-limited wireless communication channel, we formulate the following constraint optimization problem.
\begin{equation} \label{eq.SU-optimization}
    \begin{aligned}
        p^{\text{\tiny SU$_n$}}(\bm{x}_n|\bm{c})^\star = & \arg \max_{p^{\text{\tiny SU$_n$}}(\bm{x}_n|\bm{c})} & & I(\bm{\hat{x}}_n;z_n) \\
        & \text{subject to} & & I(\bm{x}_n;\bm{c}) \leq R_n.
    \end{aligned}
\end{equation}

Our formulation in (\ref{eq.SU-optimization}), aims at maximizing the mutual information between the channel output, $\bm{\hat{x}}$, and the intended semantic variable while bounding the mutual information between the encoded signal, $\bm{x}$, and the input of the SUs, $\bm{c}$. The limited rate of the corresponding channel, $R_n$, is considered to limit the number of channel uses, $d_n$, by the $n$-th SU. This is how we adopt the information bottleneck method (IBM) \cite{Tishby2000}, seeking the right balance between the inference accuracy and communication overhead using the mutual information as both an objective function and a constraint.

By employing the Lagrangian method \cite{boyd2004convex} to optimization problem (\ref{eq.SU-optimization}), we reformulate the objective function as:
\begin{equation} \label{eq.SU-optimization_lagrangian}
        \mathcal{L}^{\text{\tiny SU$_n$}} = I(\bm{\hat{x}}_n;z_n) - \lambda \,( I(\bm{x}_n;\bm{c}) - R_n).
\end{equation}

We drop the constant term, $\lambda R_n$, in (\ref{eq.SU-optimization_lagrangian}) to get the simplified equivalent objective function. Moreover, as in section \ref{subsec.CU_loss_function}, the mutual information terms in (\ref{eq.SU-optimization_lagrangian}) are generally intractable due to high-dimensional integrals. Also, following a data-driven approach, we leverage the variational approximation to form a tractable lower bound. Thus, expanding the mutual information terms and approximating the posterior distribution of the SU, \mbox{$p^{\text{\tiny SU$_n$}}(\bm{x}_n|\bm{c})$}, with NN parameterized by $\boldsymbol{\phi}_n$, yielding \mbox{$p^{\text{\tiny SU$_n$}}_{\boldsymbol{\phi}_n}(\bm{x}_n|\bm{c})$}, we end up with the following objective function:
\begin{equation} \label{eq:SU-objective}
    \begin{aligned}
        &\mathcal{L}^{\text{\tiny SU$_n$}}(\boldsymbol{\phi}_n) = I(\bm{\hat{x}}_n;z_n) - \lambda\,I(\bm{x}_n;\bm{c}) \\[0.6em]
            &\approx \mathbb{E}_{p_{\boldsymbol{\theta}}(z_n,\bm{c})} \left[\ \mathbb{E}_{p^{\text{\tiny SU}}_{\boldsymbol{\phi}_n}(\bm{x}_n|\bm{c})} \left[\,\mathbb{E}_{p(\bm{\hat{x}}_n|\bm{x}_n)} [\ \log p(z_n|\bm{\hat{x}}_n)\, ]\ \right]\ \right.\\[0.6em] 
            & \qquad \qquad \qquad \qquad - \left. \lambda\, KL(p^{\text{\tiny SU}}_{\boldsymbol{\phi}_n}(\bm{x}_n|\bm{c})\parallel p(\bm{x}_n))\right]\,,
    \end{aligned}
\end{equation}
where in (\ref{eq:SU-objective}), the outer expectation represents the effect of the pre-trained CU. The detailed derivation of the objective function above is deferred to Appendix \ref{Appendix:su_loss}. Owing to the pre-trained CU, $p_{\boldsymbol{\theta}}(z_n,\bm{c})$ is already available as:
\begin{equation} \label{eq:CU-expectation-availability}
    p_{\boldsymbol{\theta}}(z_n,\bm{c}) = \int p(z_n,\bm{S})\, p_{\boldsymbol{\theta}}(\bm{c}|\bm{S})\, d\bm{S}.
\end{equation}

The log-likelihood function appearing in the first term of the objective function (\ref{eq:SU-objective}) can be fully described when $p^{\text{\tiny SU}}_{\boldsymbol{\phi}_n}(\bm{x}_n|\bm{c})$ is available by:
\begin{equation} \label{eq:SU-log-likelihood-term}
    p(z_n|\bm{\hat{x}}_n) = \int\frac{p_{\boldsymbol{\theta}}(z_n,\bm{c})\,p^{\text{\tiny SU}}_{\boldsymbol{\phi}_n}(\bm{x}_n|\bm{c})\,p(\bm{\hat{x}}_n|\bm{x}_n)}{p(\bm{\hat{x}}_n)}\,d\bm{c}\,d\bm{x},
\end{equation}
however, same as (\ref{eq:CU-log-likelihood}), we must once more use approximations and replace the true likelihood distribution with its approximated version $p_{\boldsymbol{\psi}_n}(z_n|\bm{\hat{x}}_n)$, where $\boldsymbol{\psi}_n$ is the parameters of the corresponding NN for the $n$-th task-specific decoder. Therefore, the objective function is expressed as:
\begin{equation} \label{eq:SU-objective-final}
    \begin{aligned}
        &\mathcal{L}^{\text{\tiny SU$_n$}}(\boldsymbol{\phi}_n, \boldsymbol{\psi}_n) \approx\\[0.6em]
            &\mathbb{E}_{p_{\boldsymbol{\theta}}(z_n,\bm{c})} \left[\ \mathbb{E}_{p^{\text{\tiny SU}}_{\boldsymbol{\phi}_n}(\bm{\hat{x}}_n|\bm{c})} \left[\, \log q_{\boldsymbol{\psi}_n}(z_n|\bm{\hat{x}}_n)\, \right]\ \right.\\[0.6em] 
            & \qquad \qquad \qquad \qquad - \left. \lambda\, KL(p^{\text{\tiny SU}}_{\boldsymbol{\phi}_n}(\bm{x}_n|\bm{c})\parallel p(\bm{x}_n))\right]\,.
    \end{aligned}
\end{equation}

As we have a DNN-based implementation followed by an E2E learning fashion, improved to be effective for task-oriented communication \cite{9145068}, we emphasize performing JSCC by the SU encoders by \mbox{$p^{\text{\tiny SU}}_{\boldsymbol{\phi}_n}(\bm{\hat{x}}_n|\bm{c}) = \int p^{\text{\tiny SU}}_{\boldsymbol{\phi}_n}(\bm{x}_n|\bm{c})\, p(\bm{\hat{x}}_n|\bm{x}_n)\,d\bm{x}_n$}. This means we are taking semantic and channel statistics into account in a joint manner. In addition, we note that a detailed discussion on the approximation error analysis for the objective function in (\ref{eq:SU-objective-final}) is provided in Appendix \ref{appendix:SU_approx_error}.

For the regularization term in (\ref{eq:SU-objective-final}), where the KL divergence appears, adopting a variational marginal posterior distribution for $p(\bm{x}_n)$, which can be also called the \emph{prior distribution} of the SU's output space, is necessary. Fixing the marginal posterior distribution, or the prior, to choices such as a standard Gaussian distribution, introduced and mostly used in training variational autoencoder structures \cite{kingma2022autoencoding}, or a log-uniform distribution, which is favored for its sparsity-inducing properties, has been the most common approach taken in the literature, e.g., \cite{Shao2022,Gaussian_ep_prior1,Gaussian_ep_prior2} for adopting standard Gaussian prior and \cite{Shao2021,ep_priorloguni} for log-uniform. The primary motivation behind these choices is that they allow the KL divergence term to be computed in closed form, greatly simplifying optimization. We refer to this approach as the explicit prior (EP) method, since the prior is explicitly fixed to either a Gaussian or log-uniform distribution. However, this convenience comes at the cost of sub-optimality as the prior is restricted for mathematical tractability rather than for faithfully modeling the data. To address this limitation, we modify the loss function to incorporate density ratio estimation, enabling a more flexible and accurate approximation of the KL divergence and, in turn, a better approximation of the objective function.

\subsection{Implicit Optimal Prior Method} \label{sub:IoP}

To optimally deal with the regularization term, we modify the KL divergence in (\ref{eq:SU-objective-final}) as follows:
\begin{equation} \label{eq:SU-density-ratio-trick}
    \begin{aligned}
        & KL(p^{\text{\tiny SU}}_{\boldsymbol{\phi}_n}(\bm{x}_n|\bm{c})\parallel p(\bm{x}_n)) \\[0.6em]  & =\mathbb{E}_{p^{\text{\tiny SU}}_{\boldsymbol{\phi}_n}(\bm{x}_n|\bm{c})} \left [\ \log \frac{p^{\text{\tiny SU}}_{\boldsymbol{\phi}_n}(\bm{x}_n|\bm{c})}{p(\bm{x}_n)} \cdot \frac{q(\bm{x}_n)}{q(\bm{x}_n)} \right ]\ \\[0.6em] 
        & =KL(p^{\text{\tiny SU}}_{\boldsymbol{\phi}_n}(\bm{x}_n|\bm{c})\parallel q(\bm{x}_n)) - \mathbb{E}_{p^{\text{\tiny SU}}_{\boldsymbol{\phi}_n}(\bm{x}_n|\bm{c})} \left [\ \log \frac{p(\bm{x}_n)}{q(\bm{x}_n)} \right ]\,,
    \end{aligned}
\end{equation}
where $q(\bm{x}_n)$ is an arbitrary distribution, typically chosen as either a standard Gaussian or a log-uniform distribution, to ensure that the KL divergence can be computed in closed form. This trick, introduced in \cite{takahashi2019variational} for a variational auto-encoder, enables us to implicitly manage the prior distribution by estimating the density ratio $p(\bm{x}_n)/q(\bm{x}_n)$ and prevent fixing a distribution. \cite{takahashi2019variational} estimates the prior using a DNN-based classifier accompanied by several regularization techniques to fine-tune the estimator. In our investigations, we faced many issues while adopting the method in \cite{takahashi2019variational} to our multi-tasking SemCom framework. The issues include the convergence of the DNN-based classifier and the complexity of fine-tuning due to the existence of several regularization parameters.

To overcome these issues, we propose using classical parametric ML methods to estimate density ratios by introducing a \emph{hybrid-learning approach}. To develop our density ratio estimation for the implicit optimal prior method (IoPm) in CMT-SemCom, we follow the \emph{probabilistic classification} approach amongst other approaches of density ratio estimation \cite{Sugiyama_Suzuki_Kanamori_2012}. Using the probabilistic classification approach has advantages such as straightforward implementation and the possibility of direct use of a standard classification algorithm.

Specifically, we train a probabilistic binary classifier to distinguish between samples drawn from the arbitrary distribution, $q(\bm{x}_n)$, and samples drawn from the distribution produced by the semantic encoder, $p(\bm{x}_n)$. The key insight is that the classifier's outputs can be transformed to approximate the density ratio, which in turn allows us to compute the regularization term without requiring an explicit prior.


For this, we first sample from $q(\bm{x}_n)$ and assign labels $y=0$ to them. Next, labels $y=1$ go to samples from $p(\bm{x}_n)$ which are available using ancestral sampling \cite{bishop2006pattern} from the output of our encoder, $p^{\text{\tiny SU}}_{\boldsymbol{\phi}_n}(\mathbf{x}_n|\mathbf{c})$. Then, $p(\bm{x}_n|y)$ is defined as:
\begin{equation}
    p(\bm{x}_n|y) = 
    \begin{cases}
        p(\bm{x}_n) & y=1, \\
        q(\bm{x}_n) & y=0.
    \end{cases}
\end{equation}
Thus, our density ratio can be expressed as:
\begin{equation} \label{eq:ratio}
    \begin{aligned}
        r(\bm{x}_n) &=\frac{p(\bm{x}_n)}{q(\bm{x}_n)} = \frac{p(\bm{x}_n|y=1)}{p(\bm{x}_n|y=0)} \\[0.6em]
        & = \frac{p(y=1|\bm{x}_n)\,p(y=0)}{p(y=0|\bm{x}_n)\,p(y=1)} = \frac{p(y=1|\bm{x}_n)}{p(y=0|\bm{x}_n)},
    \end{aligned}
\end{equation}
where in (\ref{eq:ratio}), we cancel $p(y=0)$ with $p(y=1)$ since we draw an equal number of samples from both distributions. Therefore, given an estimator of the posterior probability $\hat{p}(y|\bm{x}_n)$, we can estimate the density ratio. In this work, we leverage \emph{logistic regression} (LR) classification that employs a parametric model of the following for the posterior distribution:
\begin{equation} \label{eq:LR-classifier}
    p(y|\bm{x}_n;\boldsymbol{\omega}) = \bigg ( 1 + \text{exp}\big(-y\Omega(\bm{x}_n)^{T}\boldsymbol{\omega}\big)\bigg)^{-1},
\end{equation}
\algrenewcommand\algorithmicrequire{\textbf{Input:}}
\algrenewcommand\algorithmicensure{\textbf{Output:}}
\algrenewcommand\algorithmicdo{}
\begin{algorithm}[!t]
    \caption{Density Ratio Estimation using Kernelized LR} \label{algorithm:DRE}
    \begin{algorithmic}[1]
        \Require Samples $\bm{x}_p \sim p(\mathbf{x}_n)$, samples $\bm{x}_q \sim q(\mathbf{x}_n)$, kernel bandwidth $\sigma_{k}$
        \Ensure Estimated density ratio $\hat{r}(\bm{x}_n)$
        
        \State \textbf{Step 1:} Generate labels for samples:
        \State $\mathbf{y}_{p} = \mathbf{1}_{n}$, $\mathbf{y}_{q} = -\mathbf{1}_{n}$
    
        \State \textbf{Step 2:} Combine samples and labels:
        \State $\bm{X} = [\bm{x}_{p}; \bm{x}_{q}]$
        \State $\mathbf{y} = [\mathbf{y}_{p}; \mathbf{y}_{q}]$
    
        \State \textbf{Step 3:} Compute the Gaussian kernel:
        \State $K = \exp\left(-\frac{\|\bm{X} - \bm{X}^\prime\|^2}{2\sigma^2}\right)$

        \State \textbf{Step 4:} Train logistic regression model:
        \State Fit logistic regression on $K$ with labels $\mathbf{y}$, and get $\boldsymbol{\hat{\omega}}$

        \State \textbf{Step 5:} Estimate density ratio for new data points $\bm{X}$:
        \State Compute the kernel matrix $K_{\text{new}}$ between saved $\bm{X}^\prime$ from training and the new data $\bm{X}$
        \State $\hat{r}(\bm{x_n}) = \exp\left(K_{\text{new}} \cdot \boldsymbol{\hat{\omega}} \right)$

        \State \textbf{Step 6:} Return estimated density ratio $\hat{r}(\bm{x})$
    \end{algorithmic}
\end{algorithm}
where $\Omega(\bm{x}_n)$ is a basis function and $\boldsymbol{\omega}$ is the parameter vector. Our LR model parameter is learned so that the penalized log-likelihood is maximized:
\begin{equation} \label{eq:LR-objective}
    \hat{\boldsymbol{\omega}} = \arg \max_{\boldsymbol{\omega}} \bigg[\, \sum_{k=1}^{K} \log \bigg(1 + \text{exp}\big(-y_k\Omega(\bm{x}_{n}^{(k)})^{T}\boldsymbol{\omega}\big)\bigg) + \gamma \boldsymbol{\omega}^T\boldsymbol{\omega} \bigg]\,,
\end{equation}

In (\ref{eq:LR-objective}), the term $\gamma \boldsymbol{\omega}^T\boldsymbol{\omega}$ serves as a regularization term for the LR objective function, preventing overfitting. A key advantage of the LR objective function in (\ref{eq:LR-objective}) is its convexity, which guarantees that \emph{gradient descent} (GD) methods can converge to the global optimum \cite{bertsekas1997nonlinear}. Finally, using (\ref{eq:ratio}) and (\ref{eq:LR-classifier}), our density ratio estimator (DRE) is expressed as:
\begin{equation} \label{eq:DRE}
    \hat{r}_{LR}(\bm{x}_n) = \frac{1 + \text{exp}\big(\Omega(\bm{x}_n)^{T}\boldsymbol{\hat{\omega}}\big)}{1 + \text{exp}\big(-\Omega(\bm{x}_n)^{T}\boldsymbol{\hat{\omega}}\big)} = \text{exp}\big(\Omega(\bm{x}_n)^{T}\boldsymbol{\hat{\omega}}\big).
\end{equation}

For the DRE in (\ref{eq:DRE}), we use the \emph{Gaussian kernel} for the basis function with kernel bandwidth, $\sigma_k$ as:
\begin{equation} \label{eq:kernel}
    K(\bm{x},\bm{x}^\prime) = \text{exp} \bigg( - \frac{\|\bm{x} - \bm{x}^\prime \|^2}{2\sigma_{k}^{2}} \bigg),
\end{equation}
where $\bm{x}^\prime$ denotes the stored samples of $\bm{x}_n$ from both distributions, obtained in the previous step, that we store for the next inference step of the DRE. This reflects our use of a \emph{memory-based} method \cite{bishop2006pattern} for the DRE that involves storing the samples used in training to make inferences for future data. A detailed description of the DRE procedure is provided in Algorithm \ref{algorithm:DRE}, with further discussions in Section \ref{section.Simulation}.

\begin{figure}[!t]
    \centering
    \resizebox{0.5\textwidth}{!}{


\usetikzlibrary {arrows.meta}

\definecolor{dogwoodrose}{rgb}{0.84, 0.09, 0.41}
\definecolor{dollarbill}{rgb}{0.52, 0.73, 0.4}

\def\layersep{2.0cm}
\def\channelsep{5.0cm}
\begin{tikzpicture}[shorten >=1pt,-,draw=black!30, node distance=\layersep]
    \tikzstyle{every pin edge}=[<-,shorten <=1pt]
    \tikzstyle{neuron}=[circle,fill=black!25,minimum size=12pt,inner sep=0pt]
    \tikzstyle{input neuron}=[neuron, fill=black!50];
    \tikzstyle{output neuron}=[neuron, fill=black!50];
    \tikzstyle{hidden neuron}=[neuron, fill=black!50];
    \tikzstyle{dec_in neuron}=[neuron, fill=black!50];
    \tikzstyle{dec_hidden neuron}=[neuron, fill=black!50];
    \tikzstyle{dec_out neuron}=[neuron, fill=black!50];

    \foreach \name / \y in {1,2,3,5,6,7}
        \node[input neuron] (I-\name) at (0,-\y) {};

    \draw[dotted, line width=1pt] (I-3) -- (I-5);
    \node[] (I-4) at (0,-4cm) {};
    
    \foreach \name / \y in {1,2,4,5}
        \path[yshift=-1.0cm]
            node[hidden neuron] (H-\name) at (\layersep,-\y cm) {};

    \draw[dotted, line width=1pt] (H-2) -- (H-4);
    \node[yshift=-1.0cm] (H-3) at (\layersep,-3cm) {};

    \foreach \name / \y in {1,3}
        \path[yshift=-2.0cm]
            node[output neuron] (O-\name) at ( 4cm,-\y cm) {};

    \draw[dotted, line width=1pt] (O-1) -- (O-3);
    \node[yshift=-2.0cm] (O-2) at (4cm,-2cm) {};

    \foreach \source in {1,...,7}
        \foreach \dest in {1,...,5}
            \path (I-\source) edge (H-\dest);

    \foreach \hidd in {1,...,5}
        \foreach \out in {1,...,3}
            \path (H-\hidd) edge (O-\out);

    \foreach \name / \y in {1,3}
        \path[yshift=-2.0cm]
            node[dec_in neuron] (dI-\name) at (4cm + \channelsep,-\y cm) {};

    \draw[dotted, line width=1pt] (dI-1) -- (dI-3);
    \node[yshift=-2.0cm] (dI-2) at (4cm + \channelsep,-2cm) {};

    \node[yshift=-2.0cm, dec_hidden neuron] (dH-1) at (4cm + \channelsep + \layersep,-1.3cm) {};
    \node[yshift=-2.0cm, dec_hidden neuron] (dH-3) at (4cm + \channelsep + \layersep,-2.7cm) {};
    \draw[dotted, line width=1pt] (dH-1) -- (dH-3);
    \node[yshift=-2.0cm] (dH-2) at (4cm + \channelsep + \layersep,-2cm) {};

    \node[dec_out neuron,pin={[pin distance=1cm, pin edge={-{Latex[length=3.5mm]}, black}]right:{\rotatebox{90}{\LARGE Task}}}, right of=dH-2] (O) {};

    \foreach \source in {1,...,3}
        \foreach \dest in {1,...,3}
            \path (dI-\source) edge (dH-\dest);

    \foreach \hidd in {1,...,3}
            \path (dH-\hidd) edge (O);

    \draw[orange!80,fill=orange,fill opacity=0.08,rounded corners=2]
    (-0.5,0.0) --++ (5,-2.5) --++ (0.0,-3.0) --++ (-5.0,-2.5) -- cycle;

    \draw[blue!80,fill=blue,fill opacity=0.02,rounded corners=2]
    (8.5,-2.5) --++ (5,-1.0) --++ (0.0,-1.0) --++ (-5,-1.0) -- cycle;

    \node[align=center,black] at (\layersep,0.0) {\LARGE $n$-th SU Enc.};
    \node[align=center,black] at (4cm + \channelsep + \layersep, -2.0) {\LARGE $n$-th Dec.};
    \node[left of=I-4, align=center,black, rotate=90] (input) {\LARGE Semantic Source/ CU output};

    \path[-{Latex[length=3.5mm]}, black] (input) edge (-0.5,-4);

    \draw[black,thick] (5.2cm,-5.1cm) rectangle ++(0.5cm,2.2cm);
    \foreach \i in {1,2,3} {
        \draw[black,thick] (5.2cm,-5.1cm + \i*0.55cm) -- ++(0.5cm,0);
    }

    \node[align=center, black] at (5.45,-2.5) {\LARGE $\bm{x}_n$};
    \path[-{Latex[length=3.5mm]}, black] (4.5cm,-4cm) edge (5.2cm,-4cm);

    \node[align=center, black, rotate=90] (channel) at (6.4cm,-4cm) {\Large Wireless channel};

    \draw[thick] (7.2cm,-5.1cm) rectangle ++(0.5cm,2.2cm);
    \foreach \i in {1,2,3} {
        \draw[thick] (7.2cm,-5.1cm + \i*0.55cm) -- ++(0.5cm,0);
    }

    \node[align=center, black!50] at (7.45,-2.5) {\LARGE $\bm{\hat{x}}_n$};

    \path[-{Latex[length=3.5mm]}] (7.2cm,-4cm) edge (8.5cm,-4cm);

    \draw[dashed, thick, black!70] (-1.2, 0.5) rectangle (13.8, -8.5);

    \draw[dollarbill, fill=dollarbill, fill opacity=0.6] (0.5,-9.0) rectangle ++(4.0,-2.0);
    \node[align=center, black] at (2.5,-10.0) {\huge DRE \\[0.6em] \LARGE (kernelized LR)};

    \draw[black] (8.0,-9.0) rectangle ++(4.5,-2.0);
    \node[align=center, black] at (10.25,-10.0) {\huge Loss function \\[0.6em] \LARGE (\ref{eq:SU-empirical_objective})};
    \draw[-{Latex[length=3.5mm]}, black] (8.0,-10.0) -- (6.4, -10.0) -- (6.4, -8.5);

    \draw[-{Latex[length=3.5mm]}, black] (5.45,-5.1) -- (5.45,-10.0) -- (4.5,-10.0);

    \draw[-{Latex[length=3.5mm]}, black] (0.5, -10.0) -- (-0.5, -10.0) -- (-0.5, -11.5) -- (13.5, -11.5) -- (13.5, -10.0) -- (12.5,-10.0);

\end{tikzpicture}


}
    \caption{Illustration of the proposed hybrid-learning approach for the $n$-th SU.}
    \label{fig:DRE_schematic}
\end{figure}
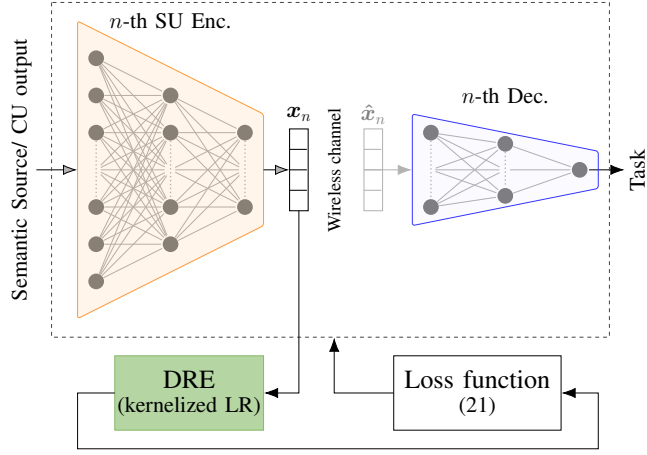

By employing the kernelized LR, we implement the IoPm within the CMT-SemCom framework to more effectively handle the regularization term, representing the communication overhead introduced by the rate-limited wireless channels. Fig. \ref{fig:DRE_schematic}, illustrates our hybrid-learning approach, which combines this classical kernelized DRE with our DNN-based semantic transmission. The output of the $n$-th SU, $\bm{x}_n$, is sampled and provided to the DRE. In the DRE unit, the samples from $q(\bm{x}_n)$ are also drawn, and the regularization is estimated by (\ref{eq:DRE}). This estimate is then used to update the SU's objective function in each training iteration. As training progresses, improved encoder outputs lead to more accurate density ratio estimates, which in turn refine the overall loss function.

Applying the discussed IoPm, the approximated objective function in (\ref{eq:SU-objective-final}) becomes:
\begin{equation} \label{eq:SU-objective-with-DRE-applied}
    \begin{aligned}
        &\mathcal{L}^{\text{\tiny SU$_n$}}(\boldsymbol{\phi}_n, \boldsymbol{\psi}_n) \approx\\[0.6em]
            &\mathbb{E}_{p_{\boldsymbol{\theta}}(z_n,\bm{c})} \left[\ \mathbb{E}_{p^{\text{\tiny SU}}_{\boldsymbol{\phi}_n}(\bm{\hat{x}}_n|\bm{c})} \left[\, \log q_{\boldsymbol{\psi}_n}(z_n|\bm{\hat{x}}_n)\, \right]\ \right.\\[0.6em] 
            &  - \left. \lambda\, \left \{ KL(p^{\text{\tiny SU}}_{\boldsymbol{\phi}_n}(\bm{x}_n|\bm{c})\parallel q(\bm{x}_n)) - \mathbb{E}_{p^{\text{\tiny SU}}_{\boldsymbol{\phi}_n}(\bm{x}_n|\bm{c})} \left [\ \log\hat{r}(\bm{x}_n) \right ]\ \right \} \right]\,.
    \end{aligned}
\end{equation}

Given that a minibatch of $\{z_{n}^{(m)},\bm{c}^{(m)}\}_{m=1}^{M_n}$ can be selected from the joint distribution $p_{\boldsymbol{\theta}}(z_n,\bm{c})$ and leveraging the Monte Carlo sampling as we did for (\ref{eq:CU-objective-empirical}), we end up with the empirical estimation of the objective function:
\begin{equation} \label{eq:SU-empirical_objective}
    \begin{aligned}
        &\mathcal{L}^{\text{\tiny SU$_n$}}(\boldsymbol{\phi}_n, \boldsymbol{\psi}_n) \approx \frac{1}{M_n} \sum^{M_n}_{m=1}  \left\{ \frac{1}{L}\sum^{L}_{l=1} \left[\,\log q_{\boldsymbol{\psi}_n}(z_n|\bm{\hat{x}}^{(m,l)}_{n})\, \right]\ \right.\\[0.6em]
        & - \left. \lambda \left[\,KL(p^{\text{\tiny SU}}_{\boldsymbol{\phi}_n}(\bm{x}^{(m)}_{n}|\bm{c})\parallel q(\bm{x}_n)) - \frac{1}{L}\sum^{L}_{l=1}\hat{r}(\bm{x}^{(m,l)}_{n}) \right]\,\right\}.
    \end{aligned}
\end{equation}

It is worth mentioning that in (\ref{eq:SU-empirical_objective}), we apply the reparameterization trick to overcome the differentiability issues in the backpropagation as previously discussed in Appendix \ref{Appendix_CU-reparameterization}. For the term, representing the corresponding channel rate, the reparameterization trick exists as \mbox{$\bm{x}^{(m,l)}_{n} = \boldsymbol{\mu}_{\bm{x}_n|\bm{c}^{m}} + \boldsymbol{\sigma}_{\bm{x}_n|\bm{c}^{m}} \odot \boldsymbol{\epsilon}^l$}. 

It is important to note that the integration of parametric DRE with DNN-based SUs is feasible in practice, particularly when using \emph{stochastic gradient descent} (SGD) as the optimizer for training the SU networks with the objective function in (\ref{eq:SU-objective-with-DRE-applied}). The use of SGD enables a key simplification by allowing us to treat the DRE as a fixed, non-trainable component during the SU training. Specifically, the term, $\mathbb{E}_{p^{\text{\tiny SU}}_{\boldsymbol{\phi}_n}(\bm{x}_n|\bm{c})} [\ \log\hat{r}(\bm{x}_n)]\,$, is not included in the backpropagation process for updating the SU parameters $\boldsymbol{\phi}_n$. This decoupling is what enables our hybrid-learning approach, where the DRE is trained separately using classical methods, while the DNN-based SUs are trained via SGD. Without this separation, the DRE would need to be differentiable and involved in gradient updates, significantly complicating the training pipeline. A detailed explanation of why this integration is compatible with SGD-based optimization is provided in Appendix \ref{appendix:SU_loss_gradients}.

\subsubsection{KL Divergence closed-Form Expression} \label{subsub:kl-closed-form}
Finally, the last step is to manage the KL divergence term in eq. (\ref{eq:SU-empirical_objective}). We assumed the Gaussian distribution for our SU encoder such that $p^{\text{\tiny SU}}_{\boldsymbol{\phi}_n}(\bm{x}_{n}|\bm{c}) \sim  \mathcal{N}(\boldsymbol{\mu}_{\bm{x}_n|\bm{c}^{m}},\boldsymbol{\sigma}_{\bm{x}_n|\bm{c}^{m}}\mathbf{I})$. Thus, the proper selection of the arbitrary prior $q(\bm{x}_n)$, e.g., standard Gaussian or log-uniform, can result in a closed-form solution for the KL term in eq. (\ref{eq:SU-empirical_objective}). In addition, since we compare our proposed method with the two most widely adopted fixed priors: the standard Gaussian and the log-uniform distribution in the EP method in Sec. \ref{section.Simulation}, we need to directly calculate the KL term in eq. (\ref{eq:SU-objective-final}) for fixed $p(\bm{x}_n)$. Here, we present the KL divergence calculation for both cases.

Standard Gaussian Prior: When the arbitrary distribution $q(\bm{x}_n)$ in eq. (\ref{eq:SU-empirical_objective}) (or, equivalently, the fixed prior in the EP method $p(\bm{x}_n)$ in eq. (\ref{eq:SU-objective-final})) is chosen as the standard Gaussian, the KL divergence reduces to the well-known closed-form expression between two Gaussian distributions:
\begin{equation} \label{eq:KL-closed-form-expression}
    \begin{aligned}
        &KL\left(p^{\text{\tiny SU}}_{\boldsymbol{\phi}_n}(\bm{x}_{n}|\bm{c})\parallel q(\bm{x}_n)\right) \\[0.6em] &= \frac{1}{2} \left (\,\log\frac{1}{\sigma^{2}_{x_{n}|\bm{c}}} + \sigma^{2}_{x_{n}|\bm{c}} + \mu^{2}_{x_{n}|\bm{c}} -1 \right)\,.
    \end{aligned}    
\end{equation}

We note that this results from computing the KL divergence between the output distribution of each SU encoder, $p^{\text{\tiny SU}}_{\boldsymbol{\phi}_n}(\bm{x}_n|\bm{c})$ and its corresponding prior $q(\bm{x}_n)$ (or, equivalently, $p(\bm{x}_n)$ in the EP method). In other words, the SU objective ($\mathcal{L}^{\text{\tiny SU$_n$}}(\boldsymbol{\phi}_n, \boldsymbol{\psi}_n)$) is per SU and is calculated for the $n$-th SU.

Log-Uniform Prior: Alternately, when $q(\bm{x}_n)$ (or equivalently, $p(\bm{x}_n)$ in the EP method) is selected as a log-uniform distribution, the KL divergence can also be approximately expressed in closed form by taking advantage of the results of \cite{kingma2015variational,molchanov2017variational} as follows:
\begin{equation} \label{eq:KL-loguni}
\begin{aligned}
&KL\left(p^{\text{\tiny SU}}_{\boldsymbol{\phi}_n}(x^{(i)}_{n}|\bm{c})\parallel q(x^{(i)}_n)\right) 
\\[0.6em]&= \tfrac{1}{2}\log \alpha_i 
   - \mathbb{E}_{\epsilon \sim \mathcal{N}(1,\alpha_i)}\!\big[\log|\epsilon|\big] + C \\[0.6em]
&\approx k_1 \,\sigma\!\left(k_2 + k_3 \log \alpha_i\right) 
   - \tfrac{1}{2} \log\!\left(1+\alpha_i^{-1}\right) + C,
\end{aligned}
\end{equation}
where 
\[
\alpha_i = \frac{\sigma^{2}_{x^{(i)}_{n}|\bm{c}}}{x^{(i)^2}_n}, 
\quad k_1 = 0.63576, 
\quad k_2 = 1.87320, 
\quad k_3 = 1.48695.
\]
and $C$ is a constant. Besides, $x^{(i)}_n$ is the $i$-th dimension in $\bm{x}_n$, and $\sigma(\cdot)$ denotes the sigmoid function.

Consequently, the empirical approximation of the objective function in (\ref{eq:SU-objective-final}) is calculated as above. The training procedure for the $n$-th SU adopting the standard Gaussian for the arbitrary prior is described in Algorithm \ref{algorithm:SU_training}.

\algrenewcommand\algorithmicrequire{\textbf{Input:}}
\algrenewcommand\algorithmicensure{\textbf{Output:}}
\begin{algorithm}[!t]
    \caption{Training the $n$-th SU Encoder and Decoder.} \label{algorithm:SU_training}
    \begin{algorithmic}[1]
        \Require \text{Preprocessed training dataset:} $\{\bm{S}^{(m)},z^{(m)}_{1},\dots, z^{(m)}_{N}\}_{m\in M}$, optimized parameters $\boldsymbol{\theta}$, number of iterations $T$, batch sizes $M_n$
        \Ensure The trained parameters $\boldsymbol{\phi}_n, \boldsymbol{\psi}_n,$ and $\, \boldsymbol{\omega}$
        \For{epoch $t=1$ to $T$}
        \State \text{Randomly select a minibatch $\{(\bm{S}^{(m)},z_{n}^{(m)}) \}_{m=1}^{M_n}$}
        \State \parbox[t]{\dimexpr\linewidth-\algorithmicindent}{Extract $\{\bm{c}^m\}_{m=1}^{M_n}$ from the learned $p^{\text{\tiny CU}}_{\boldsymbol{\theta}}(\bm{c}|\bm{S})$}
        \State \parbox[t]{\dimexpr\linewidth-\algorithmicindent}{compute the mean vector $\{\boldsymbol{\mu}_{\bm{x}_n|\bm{c}^{m}}\}_{m=1}^{M_n}$ and the standard deviation vector $\{\boldsymbol{\sigma}_{\bm{c}|\bm{S}^{m}}\}_{m=1}^{M_n}$}
        \For{$m=1$ to $M_n$}
            \State \text{Sample the $\{\boldsymbol{\epsilon}^{(m,l)}\}_{l=1}^{L} \sim \mathcal{N}(\mathbf{0}, \mathbf{I})$}
            \State \text{Compute $\bm{x}^{(m,l)}_{n} = \boldsymbol{\mu}_{\bm{x}_n|\bm{c}^{m}} + \boldsymbol{\sigma}_{\bm{x}_n|\bm{c}^{m}} \odot \boldsymbol{\epsilon}^{(m,l)}$}
        \EndFor    
        \State \text{Compute the log-likelihood $\log q_{\boldsymbol{\psi}_n}(z_n|\bm{\hat{x}}^{(m,l)}_{n})$}
        \State Compute the density ratio based on Algorithm \ref{algorithm:DRE}
        \State Compute the gradients of $\mathcal{L}^{\text{\tiny SU$_n$}}$ in (\ref{eq:SU-empirical_objective})
        \State Update parameters $\boldsymbol{\phi}_n$ and $\boldsymbol{\psi}_n$
        \State Compute the total loss value $\mathcal{L}^{\text{\tiny SU$_n$}}$ based on (\ref{eq:SU-empirical_objective}).
        \EndFor 
    \end{algorithmic}
\end{algorithm}

\section{Simulation Results} \label{section.Simulation}

To evaluate the effectiveness of our proposed CMT-SemCom design over rate-limited wireless channels using the hybrid-learning framework, we consider two representative tasks in our multi-label tasks paradigm: binary and categorical classification. These correspond to two different semantic variables, modeled as $z_1 \sim Bernoulli$ and $z_2 \sim Multinomial$. We begin by assessing the accuracy of our proposed density ratio estimator. Then, we present the overall performance of the CMT-SemCom across various datasets. In addition, we examine system behavior under different levels of channel constraint. Finally, we compare the proposed IoPm with two widely used baselines that follow explicit fixed prior (EP) method (standard Gaussian and log-uniform prior).  \footnote{The simulation code of this paper is available at \url{https://github.com/ant-uni-bremen/CMT-SemCom_IoPm}}.

\subsection{Simulation Setup}
We examine the proposed framework across two widely adopted datasets in semantic/task-oriented communication \cite{ImageRecoveryClassification,LatentDiffusion,ExplainableSemantic,RobustInformationBottleneck,Multi-DeviceTask-Oriented,Shao2021,Shao2022}. The MNIST dataset of handwritten digits \cite{deng2012mnist}, contains 60,000 images for the training set and 10,000 samples for the test set. Moreover, the CIFAR-10 dataset \cite{cifar} consists of 60000, $32\times32$ color images in 10 classes, with 6000 images per class. There are 50000 training images and 10000 test images. For a specific number of tasks denoted by $T$, we shape our semantic source as $\{ \bm{S}^{(m)}, z^{(m)}_1, ..., z^{(m)}_T \} ^{M}_{m=1}$. The implemented DNN structure and the specification of the parameterized DRE are listed in Table \ref{table_implementation} and Table \ref{table_implementation_dre}, respectively. The specifications are found heuristically such that the performance is maximized.

\begin{table}[!t]
    \centering
    \caption{Encoder-decoder NN architecture for the proposed CMT-SemCom.}
    \label{table_implementation}
    \begin{subtable}{\columnwidth}
        \centering
        \caption{NN structure for the MNIST dataset}
        \begin{tabular}{c|l|p{0.4\columnwidth}}
            \hline
            & \textbf{Layer} & \textbf{Properties} \\
            \hline
            \multirow{4}{*}{\textbf{CU}} & Dense & size: 256, activation: ReLU \\
            & Dense & size: 256, activation: ReLU \\
            & Dense ($\boldsymbol{\mu}_{\bm{c}}$) & size: 128, activation: Linear \\
            & Dense ($\boldsymbol{\sigma}_{\bm{c}}$) & size: 128, activation: Linear \\
            \hline
            \multirow{3}{*}{\textbf{Dec$_{\text{aux}}$}} & Dense & size: 128, activation: ReLU \\
            & Dense (T$_1$) & size: 1, activation: Sigmoid \\
            & Dense (T$_2$) & size: 10, activation: Softmax \\
            \hline
            \multirow{4}{*}{\textbf{SU}} & Dense & size: 64, activation: ReLU \\
            & Dense & size: 64, activation: ReLU \\
            & Dense ($\boldsymbol{\mu}_{\bm{x}_n}$) & size: 32, activation: Tanh \\
            & Dense ($\boldsymbol{\sigma}_{\bm{x}_n}$) & size: 32, activation: Sigmoid \\
            \hline
            \multirow{3}{*}{\textbf{Dec}} & Dense & size: 32, activation: ReLU \\
            & Dense (T$_1$) & size: 1, activation: Sigmoid \\
            & Dense (T$_2$) & size: 10, activation: Softmax
        \end{tabular}
    \end{subtable}
    \begin{subtable}{\columnwidth}
        \centering
        \caption{NN structure for the CIFAR-10 dataset}
        \begin{tabular}{c|l|p{0.4\columnwidth}}
            \hline
            & \textbf{Layer} & \textbf{Properties} \\
            \hline
            \multirow{13}{*}{\textbf{CU}} & Conv2D & filter size: 32, kernel size: (8,8), activation: ReLU \\
            & Conv2D & filter size: 32, kernel size: (8,8), activation: ReLU \\
            & MaxPooling2D & pool size: (2,2)\\
            & Dropout & dropout rate: 0.1 \\
            & Conv2D & filter size: 32, kernel size: (8,8), activation: ReLU \\
            & MaxPooling2D & pool size: (2,2)\\
            & Dropout & dropout rate: 0.2 \\
            & Conv2D & filter size: 32, kernel size: (8,8), activation: ReLU \\
            & MaxPooling2D & pool size: (2,2)\\
            & Dropout & dropout rate: 0.2 \\
            & Flatten & - \\
            & Dense ($\boldsymbol{\mu}_{\bm{c}}$) & size: 256, activation: Linear\\
            & Dense ($\boldsymbol{\sigma}_{\bm{c}}$) & size: 256, activation: Linear\\
            \hline
            \multirow{5}{*}{\textbf{Dec$_{\text{aux}}$}} & Dense & size: 256, activation: ReLU \\
            & Dense & size: 128, activation: ReLU \\
            & Dropout & dropout rate: 0.2 \\
            & Dense (T$_1$) & size: 1, activation: Sigmoid \\
            & Dense (T$_2$) & size: 10, activation: Softmax \\
            \hline
            \multirow{4}{*}{\textbf{SU}} & Dense & size: 256, activation: ReLU \\
            & Dense & size: 256, activation: ReLU \\
            & Dense ($\boldsymbol{\mu}_{\bm{x}_n}$) & size: 128, activation: Tanh \\
            & Dense ($\boldsymbol{\sigma}_{\bm{x}_n}$) & size: 128, activation: Sigmoid \\
            \hline
            \multirow{3}{*}{\textbf{Dec}} & Dense & size: 128, activation: ReLU \\
            & Dense (T$_1$) & size: 1, activation: Sigmoid \\
            & Dense (T$_2$) & size: 10, activation: Softmax
        \end{tabular}
    \end{subtable}
\end{table}

\subsection{DRE Performance}
To implement the IoPm using the density ratio trick, we initially explored a DNN-based approach for the DRE, motivated by the generalization capabilities of DNNs. However, within the context of our CMT-SemCom framework, we encountered various challenges related to convergence and hyperparameter tuning. For simple cases like a single variational autoencoder, i.e.,\cite{takahashi2019variational}, many techniques such as dropout, dynamic binarization, early stopping, etc., are employed together to fine-tune the estimator. However, these strategies failed to stabilize training in our more complex multi-task setting.

\begin{table}[!t]
        \centering
        \caption{Specification of the parametric kernelized DRE.}
        \label{table_implementation_dre}
        \begin{tabular}{c|l|l}
            \hline
            & \textbf{Component} & \textbf{Description} \\
            \hline
            & Model & Logistic Regression \\
            & Kernel & Radial Basis Function kernel \\
            & Training & Quasi-Newton Method \\
            \hline
            \multirow{3}{*}{MNIST} & Kernel BW & $\sigma_k = 1.9$ \\
            & Regularization & $\gamma = 1.5$ \\
            & Sample size & 2000 \\
            \hline
            \multirow{3}{*}{CIFAR-10} & Kernel BW & $\sigma_k = 5.0$ \\
            & Regularization & $\gamma = 2.0$\\
            & Sample size & 4000 
        \end{tabular}
\end{table}

As a result, we turned to a classical parametric ML method, which offers a simpler and more reliable implementation. This approach not only avoids the instability of DNN training but also provides the potential for optimal estimation under correct model specification, making it well-suited for our hybrid-learning framework.

\begin{figure}[H]
    \centering
    \resizebox{0.40\textwidth}{!}{\input{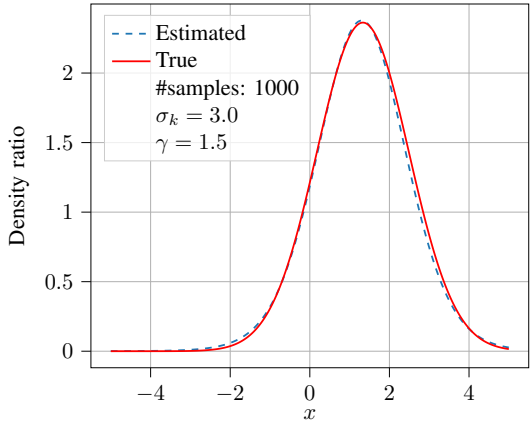}}
    \caption{Performance of the proposed DRE for the scalar data.}
    \label{fig:1D-DRE}
\end{figure}

We first evaluate the behavior of the DRE in a simple one-dimensional setting. As shown in Fig. \ref{fig:1D-DRE}, the estimator accurately captures the density ratio between two univariate Gaussian distributions, $x_1 \sim \mathcal{N}(0, 1)$ and $x_2 \sim \mathcal{N}(1, 2)$. The performance changes depending on the DRE's specification, i.e., sample size, kernel, etc., and the specification used in our evaluations is included in the figures.
 
\begin{figure*}[t!]
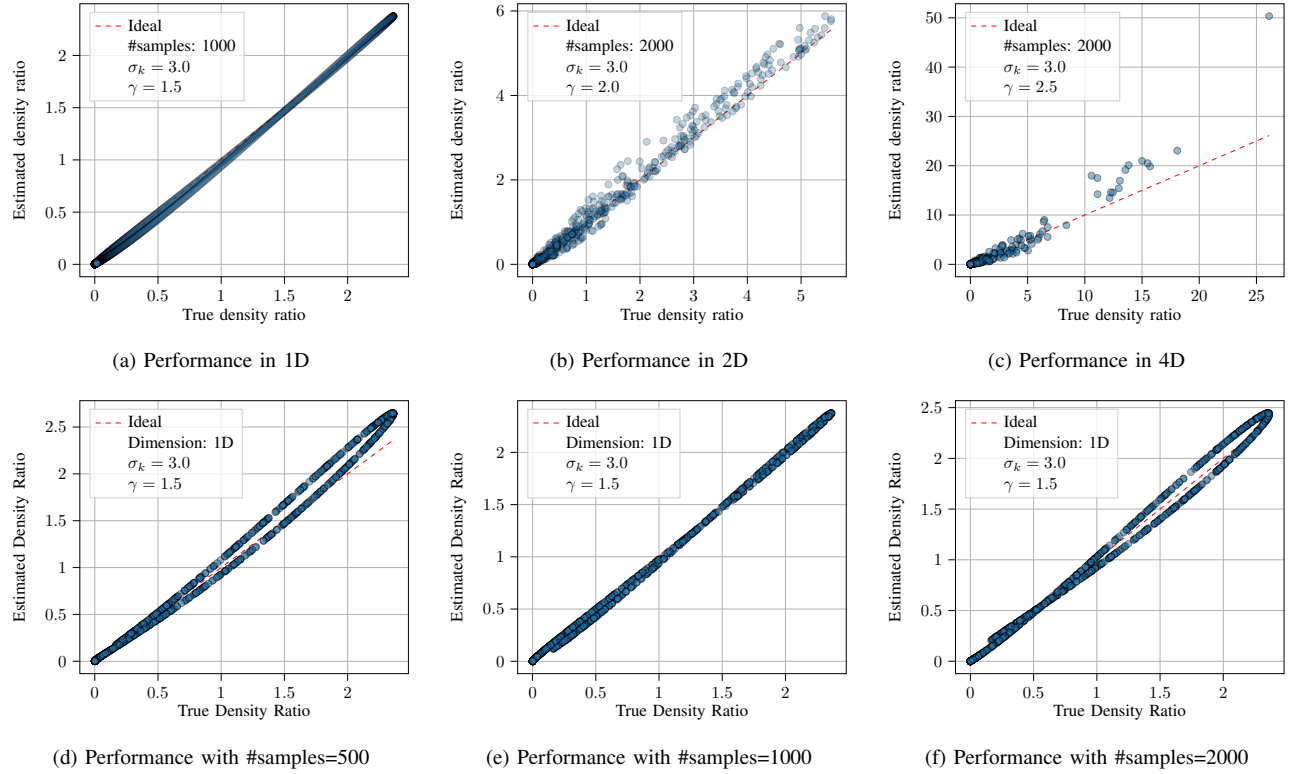

    \centering
    \begin{subfigure}[b]{0.32\textwidth}
        \centering
        \resizebox{0.95\textwidth}{!}{\input{Figures/scatter/DRE_1D_scatter.tikz}}
        \caption{Performance in 1D}
        \label{fig:scatter-DRE_a}
    \end{subfigure}
    \begin{subfigure}[b]{0.32\textwidth}
        \centering
        \resizebox{0.95\textwidth}{!}{\input{Figures/scatter/DRE_2D_scatter_s2.tikz}}
        \caption{Performance in 2D}
        \label{fig:scatter-DRE_b}
    \end{subfigure}
    \begin{subfigure}[b]{0.32\textwidth}
        \centering
        \resizebox{0.95\textwidth}{!}{\input{Figures/scatter/DRE_4D_scatter_s3.tikz}}
        \caption{Performance in 4D}
        \label{fig:scatter-DRE_c}
    \end{subfigure}
    \vspace{0.5em}
    \begin{subfigure}[b]{0.32\textwidth}
        \centering
        \resizebox{0.95\textwidth}{!}{\input{Figures/scatter/DRE_1D_Sample500.tikz}}
        \caption{Performance with \#samples=500}
        \label{fig:scatter-DRE_d}
    \end{subfigure}
    \begin{subfigure}[b]{0.32\textwidth}
        \centering
        \resizebox{0.95\textwidth}{!}{\input{Figures/scatter/DRE_1D_Sample1000.tikz}}
        \caption{Performance with \#samples=1000}
        \label{fig:scatter-DRE_e}
    \end{subfigure}
    \begin{subfigure}[b]{0.32\textwidth}
        \centering
        \resizebox{0.95\textwidth}{!}{\input{Figures/scatter/DRE_1D_Sample2000.tikz}}
        \caption{Performance with \#samples=2000}
        \label{fig:scatter-DRE_f}
    \end{subfigure}
    \caption{Performance of the proposed DRE for different specifications of data dimensions and number of samples.}
    \label{fig:scatter-DRE}
\end{figure*}

To inspect how dimensionality affects the performance of the proposed DRE, we extend this analysis to multivariate cases with increasing dimensions. Figs.~\ref{fig:scatter-DRE} (\subref{fig:scatter-DRE_a},
\subref{fig:scatter-DRE_b}, \subref{fig:scatter-DRE_c}) present scatter plots of the estimated density ratios in 1D, 2D, and 4D. The evaluations in these figures, are for estimating the density ratio of two multivariate Gaussian distribution. For instance, for the 4D case the distributions are $\mathbf{x}_1 \sim \mathcal{N}(\boldsymbol{\mu}_1, \boldsymbol{\Sigma}_1)$, where $\boldsymbol{\mu}_1 = \begin{bmatrix} 1 & 1 & 1 & 1 \end{bmatrix}^\top$ and $\boldsymbol{\Sigma}_1 = \mathbf{I}_{d=4}$, and $\mathbf{x}_2 \sim \mathcal{N}(\boldsymbol{\mu}_2, \boldsymbol{\Sigma}_2)$, where $\boldsymbol{\mu}_2 = \begin{bmatrix} 0 & 0 & 0 & 0 \end{bmatrix}^\top$ and $\boldsymbol{\Sigma}_2 = 4\cdot\mathbf{I}_{d=4}$. We observe that as the dimension increases, the accuracy of the DRE decreases. 

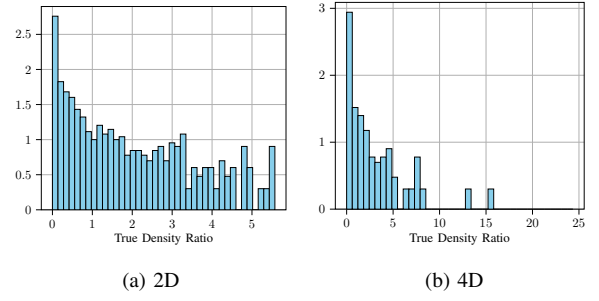
\begin{figure}[t]
    \centering
    \begin{subfigure}[b]{0.45\linewidth}
        \centering
        \resizebox{0.95\textwidth}{!}{
\begin{tikzpicture}

\definecolor{darkgray176}{RGB}{176,176,176}
\definecolor{skyblue}{RGB}{135,206,235}

\begin{axis}[
tick align=outside,
tick pos=left,
x grid style={darkgray176},
xlabel={True Density Ratio},
xmajorgrids,
xmin=-0.278703947885374, xmax=5.85278308218507,
xtick style={color=black},
y grid style={darkgray176},
ymajorgrids,
ymin=0, ymax=2.89685748701787,
ytick style={color=black}
]
\draw[draw=black,fill=skyblue] (axis cs:8.02691852235658e-09,0) rectangle (axis cs:0.139351985983065,2.75891189239797);
\draw[draw=black,fill=skyblue] (axis cs:0.139351985983065,0) rectangle (axis cs:0.278703963939211,1.82607480270083);
\draw[draw=black,fill=skyblue] (axis cs:0.278703963939211,0) rectangle (axis cs:0.418055941895358,1.68124123737559);
\draw[draw=black,fill=skyblue] (axis cs:0.418055941895358,0) rectangle (axis cs:0.557407919851504,1.60205999132796);
\draw[draw=black,fill=skyblue] (axis cs:0.557407919851504,0) rectangle (axis cs:0.69675989780765,1.43136376415899);
\draw[draw=black,fill=skyblue] (axis cs:0.69675989780765,0) rectangle (axis cs:0.836111875763797,1.32221929473392);
\draw[draw=black,fill=skyblue] (axis cs:0.836111875763797,0) rectangle (axis cs:0.975463853719943,1.11394335230684);
\draw[draw=black,fill=skyblue] (axis cs:0.975463853719943,0) rectangle (axis cs:1.11481583167609,1);
\draw[draw=black,fill=skyblue] (axis cs:1.11481583167609,0) rectangle (axis cs:1.25416780963224,1.20411998265592);
\draw[draw=black,fill=skyblue] (axis cs:1.25416780963224,0) rectangle (axis cs:1.39351978758838,1.07918124604762);
\draw[draw=black,fill=skyblue] (axis cs:1.39351978758838,0) rectangle (axis cs:1.53287176554453,1.14612803567824);
\draw[draw=black,fill=skyblue] (axis cs:1.53287176554453,0) rectangle (axis cs:1.67222374350067,1);
\draw[draw=black,fill=skyblue] (axis cs:1.67222374350067,0) rectangle (axis cs:1.81157572145682,1.04139268515822);
\draw[draw=black,fill=skyblue] (axis cs:1.81157572145682,0) rectangle (axis cs:1.95092769941297,0.778151250383644);
\draw[draw=black,fill=skyblue] (axis cs:1.95092769941297,0) rectangle (axis cs:2.09027967736911,0.845098040014257);
\draw[draw=black,fill=skyblue] (axis cs:2.09027967736911,0) rectangle (axis cs:2.22963165532526,0.845098040014257);
\draw[draw=black,fill=skyblue] (axis cs:2.22963165532526,0) rectangle (axis cs:2.36898363328141,0.778151250383644);
\draw[draw=black,fill=skyblue] (axis cs:2.36898363328141,0) rectangle (axis cs:2.50833561123755,0.698970004336019);
\draw[draw=black,fill=skyblue] (axis cs:2.50833561123755,0) rectangle (axis cs:2.6476875891937,0.845098040014257);
\draw[draw=black,fill=skyblue] (axis cs:2.6476875891937,0) rectangle (axis cs:2.78703956714985,0.903089986991944);
\draw[draw=black,fill=skyblue] (axis cs:2.78703956714985,0) rectangle (axis cs:2.92639154510599,0.698970004336019);
\draw[draw=black,fill=skyblue] (axis cs:2.92639154510599,0) rectangle (axis cs:3.06574352306214,0.954242509439325);
\draw[draw=black,fill=skyblue] (axis cs:3.06574352306214,0) rectangle (axis cs:3.20509550101828,0.903089986991944);
\draw[draw=black,fill=skyblue] (axis cs:3.20509550101828,0) rectangle (axis cs:3.34444747897443,1.07918124604762);
\draw[draw=black,fill=skyblue] (axis cs:3.34444747897443,0) rectangle (axis cs:3.48379945693058,0.301029995663981);
\draw[draw=black,fill=skyblue] (axis cs:3.48379945693058,0) rectangle (axis cs:3.62315143488672,0.602059991327962);
\draw[draw=black,fill=skyblue] (axis cs:3.62315143488672,0) rectangle (axis cs:3.76250341284287,0.477121254719662);
\draw[draw=black,fill=skyblue] (axis cs:3.76250341284287,0) rectangle (axis cs:3.90185539079902,0.602059991327962);
\draw[draw=black,fill=skyblue] (axis cs:3.90185539079902,0) rectangle (axis cs:4.04120736875516,0.602059991327962);
\draw[draw=black,fill=skyblue] (axis cs:4.04120736875516,0) rectangle (axis cs:4.18055934671131,0.301029995663981);
\draw[draw=black,fill=skyblue] (axis cs:4.18055934671131,0) rectangle (axis cs:4.31991132466745,0.698970004336019);
\draw[draw=black,fill=skyblue] (axis cs:4.31991132466745,0) rectangle (axis cs:4.4592633026236,0.477121254719662);
\draw[draw=black,fill=skyblue] (axis cs:4.4592633026236,0) rectangle (axis cs:4.59861528057975,0.602059991327962);
\draw[draw=black,fill=skyblue] (axis cs:4.59861528057975,0) rectangle (axis cs:4.73796725853589,0);
\draw[draw=black,fill=skyblue] (axis cs:4.73796725853589,0) rectangle (axis cs:4.87731923649204,0.903089986991944);
\draw[draw=black,fill=skyblue] (axis cs:4.87731923649204,0) rectangle (axis cs:5.01667121444819,0.602059991327962);
\draw[draw=black,fill=skyblue] (axis cs:5.01667121444819,0) rectangle (axis cs:5.15602319240433,0);
\draw[draw=black,fill=skyblue] (axis cs:5.15602319240433,0) rectangle (axis cs:5.29537517036048,0.301029995663981);
\draw[draw=black,fill=skyblue] (axis cs:5.29537517036048,0) rectangle (axis cs:5.43472714831663,0.301029995663981);
\draw[draw=black,fill=skyblue] (axis cs:5.43472714831663,0) rectangle (axis cs:5.57407912627277,0.903089986991944);
\end{axis}

\end{tikzpicture}}
        \caption{2D}
    \end{subfigure}    
    \begin{subfigure}[b]{0.45\linewidth}
        \centering
        \resizebox{0.95\textwidth}{!}{
\begin{tikzpicture}

\definecolor{darkgray176}{RGB}{176,176,176}
\definecolor{skyblue}{RGB}{135,206,235}

\begin{axis}[
tick align=outside,
tick pos=left,
x grid style={darkgray176},
xlabel={True Density Ratio},
xmajorgrids,
xmin=-1.21729925928987, xmax=25.5632844450874,
xtick style={color=black},
y grid style={darkgray176},
ymajorgrids,
ymin=0, ymax=3.08858700426612,
ytick style={color=black}
]
\draw[draw=black,fill=skyblue] (axis cs:8.53376765455971e-15,0) rectangle (axis cs:0.608649629644947,2.9415114326344);
\draw[draw=black,fill=skyblue] (axis cs:0.608649629644947,0) rectangle (axis cs:1.21729925928988,1.51851393987789);
\draw[draw=black,fill=skyblue] (axis cs:1.21729925928988,0) rectangle (axis cs:1.82594888893482,1.39794000867204);
\draw[draw=black,fill=skyblue] (axis cs:1.82594888893482,0) rectangle (axis cs:2.43459851857976,1.17609125905568);
\draw[draw=black,fill=skyblue] (axis cs:2.43459851857976,0) rectangle (axis cs:3.0432481482247,0.778151250383644);
\draw[draw=black,fill=skyblue] (axis cs:3.0432481482247,0) rectangle (axis cs:3.65189777786964,0.698970004336019);
\draw[draw=black,fill=skyblue] (axis cs:3.65189777786964,0) rectangle (axis cs:4.26054740751457,0.778151250383644);
\draw[draw=black,fill=skyblue] (axis cs:4.26054740751457,0) rectangle (axis cs:4.86919703715951,0.903089986991944);
\draw[draw=black,fill=skyblue] (axis cs:4.86919703715951,0) rectangle (axis cs:5.47784666680445,0.477121254719662);
\draw[draw=black,fill=skyblue] (axis cs:5.47784666680445,0) rectangle (axis cs:6.08649629644939,0);
\draw[draw=black,fill=skyblue] (axis cs:6.08649629644939,0) rectangle (axis cs:6.69514592609433,0.301029995663981);
\draw[draw=black,fill=skyblue] (axis cs:6.69514592609433,0) rectangle (axis cs:7.30379555573926,0.301029995663981);
\draw[draw=black,fill=skyblue] (axis cs:7.30379555573926,0) rectangle (axis cs:7.9124451853842,0.778151250383644);
\draw[draw=black,fill=skyblue] (axis cs:7.9124451853842,0) rectangle (axis cs:8.52109481502914,0.301029995663981);
\draw[draw=black,fill=skyblue] (axis cs:8.52109481502914,0) rectangle (axis cs:9.12974444467408,0);
\draw[draw=black,fill=skyblue] (axis cs:9.12974444467408,0) rectangle (axis cs:9.73839407431902,0);
\draw[draw=black,fill=skyblue] (axis cs:9.73839407431902,0) rectangle (axis cs:10.347043703964,0);
\draw[draw=black,fill=skyblue] (axis cs:10.347043703964,0) rectangle (axis cs:10.9556933336089,0);
\draw[draw=black,fill=skyblue] (axis cs:10.9556933336089,0) rectangle (axis cs:11.5643429632538,0);
\draw[draw=black,fill=skyblue] (axis cs:11.5643429632538,0) rectangle (axis cs:12.1729925928988,0);
\draw[draw=black,fill=skyblue] (axis cs:12.1729925928988,0) rectangle (axis cs:12.7816422225437,0);
\draw[draw=black,fill=skyblue] (axis cs:12.7816422225437,0) rectangle (axis cs:13.3902918521886,0.301029995663981);
\draw[draw=black,fill=skyblue] (axis cs:13.3902918521886,0) rectangle (axis cs:13.9989414818336,0);
\draw[draw=black,fill=skyblue] (axis cs:13.9989414818336,0) rectangle (axis cs:14.6075911114785,0);
\draw[draw=black,fill=skyblue] (axis cs:14.6075911114785,0) rectangle (axis cs:15.2162407411235,0);
\draw[draw=black,fill=skyblue] (axis cs:15.2162407411235,0) rectangle (axis cs:15.8248903707684,0.301029995663981);
\draw[draw=black,fill=skyblue] (axis cs:15.8248903707684,0) rectangle (axis cs:16.4335400004133,0);
\draw[draw=black,fill=skyblue] (axis cs:16.4335400004133,0) rectangle (axis cs:17.0421896300583,0);
\draw[draw=black,fill=skyblue] (axis cs:17.0421896300583,0) rectangle (axis cs:17.6508392597032,0);
\draw[draw=black,fill=skyblue] (axis cs:17.6508392597032,0) rectangle (axis cs:18.2594888893481,0);
\draw[draw=black,fill=skyblue] (axis cs:18.2594888893481,0) rectangle (axis cs:18.8681385189931,0);
\draw[draw=black,fill=skyblue] (axis cs:18.8681385189931,0) rectangle (axis cs:19.476788148638,0);
\draw[draw=black,fill=skyblue] (axis cs:19.476788148638,0) rectangle (axis cs:20.085437778283,0);
\draw[draw=black,fill=skyblue] (axis cs:20.085437778283,0) rectangle (axis cs:20.6940874079279,0);
\draw[draw=black,fill=skyblue] (axis cs:20.6940874079279,0) rectangle (axis cs:21.3027370375728,0);
\draw[draw=black,fill=skyblue] (axis cs:21.3027370375728,0) rectangle (axis cs:21.9113866672178,0);
\draw[draw=black,fill=skyblue] (axis cs:21.9113866672178,0) rectangle (axis cs:22.5200362968627,0);
\draw[draw=black,fill=skyblue] (axis cs:22.5200362968627,0) rectangle (axis cs:23.1286859265077,0);
\draw[draw=black,fill=skyblue] (axis cs:23.1286859265077,0) rectangle (axis cs:23.7373355561526,0);
\draw[draw=black,fill=skyblue] (axis cs:23.7373355561526,0) rectangle (axis cs:24.3459851857975,0);
\end{axis}

\end{tikzpicture}}
        \caption{4D}
    \end{subfigure}
    \caption{The histogram of the true density ratios for different data dimensions.}
    \label{fig:histogram-DRE}
\end{figure}

This decline in performance is further illustrated in Fig. \ref{fig:histogram-DRE}, which shows the histograms of the true density ratios for 2D and 4D cases. We observe that as the dimension increases, the values of the density ratios tend to concentrate around lower values. This concentration makes it difficult for the estimator to distinguish between regions of high and low density, especially under limited sampling. 

Further, we evaluate the impact of varying sample sizes on the DRE performance. To investigate the effect of sample size, we evaluate the DRE in the 1D case and keep all other specifications the same across different sample sizes. Figs.~\ref{fig:scatter-DRE} (\subref{fig:scatter-DRE_a}, \subref{fig:scatter-DRE_b}, \subref{fig:scatter-DRE_c}) illustrate this impact for sample sizes of 500, 1000, and 2000. We observe that the DRE benefits from increased sample sizes up to a point but it degrades with more samples due to the fact that the kernel becomes noise sensitive, and can have memory expansion as well. For investigation on the impact of other DRE specifications, e.g., kernel bandwidth, we invite the interested readers to check our published simulation codes available at \url{https://github.com/ant-uni-bremen/CMT-SemCom_IoPm}.

We note that for our further experimental evaluations, we employed a grid search with cross-validation to select the best combination for the DRE specifications, and these specifications are listed in Table \ref{table_implementation} for different datasets. Moreover, we examined other kernels for the basis function, such as the \emph{Polynomial kernel} and the \emph{Sigmoid kernel}, to inspect their ability to capture the complex interactions in high dimensions, but the Gaussian kernel still had the best performance. 

\subsection{Impact of the IoPm on Cooperative Multi-Tasking}
We evaluate the effectiveness of the proposed rate-limited CMT-SemCom enabled by IoPm by measuring task execution error rates. Specifically, we compare two scenarios:
\begin{itemize}
    \item w.CU (with CU): Both SUs cooperate through the CU to execute their tasks.  \item w.o.CU (without CU): CU is omitted and each SU execute its task independently, using the semantic source directly as input without any cooperation.
\end{itemize}

\begin{figure}[t]
    \centering
    \begin{subfigure}{0.95\linewidth}
        \centering
        \resizebox{0.95\textwidth}{!}{\input{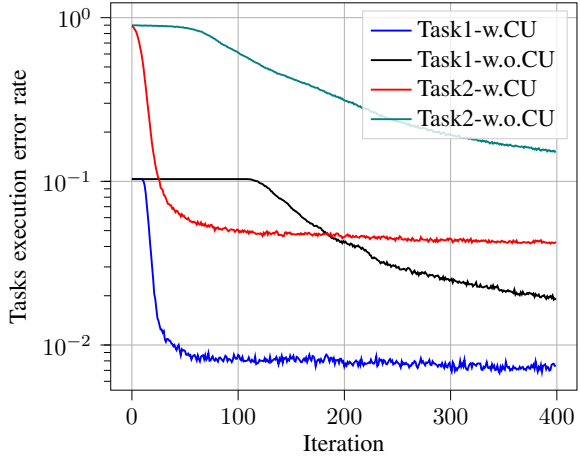}}
        \caption{Performance on MNIST dataset.}
        \label{fig:wwo_cu_mnist}
    \end{subfigure}  
    \vspace{0.4em}
    \begin{subfigure}{0.95\linewidth}
        \centering
        \resizebox{0.95\textwidth}{!}{\input{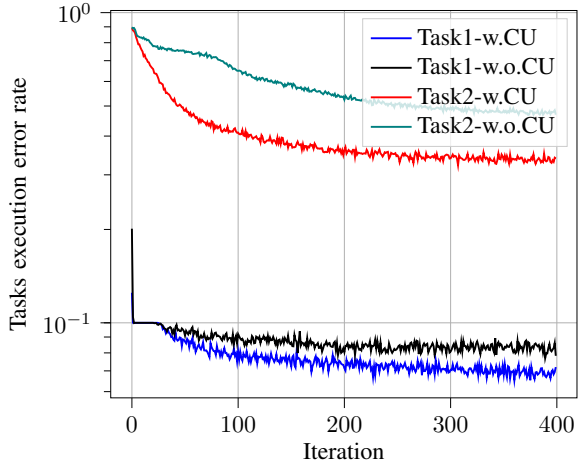}}
        \caption{Performance on CIFAR dataset.}
        \label{fig:wwo_cu_cifar}
    \end{subfigure}
    \caption{Performance of the CMT-SemCom under the hybrid-learning IoPm.}
\end{figure}

Fig. \ref{fig:wwo_cu_mnist} presents the comparison for the MNIST dataset. The results clearly show that the cooperative processing case (w.CU), CMT-SemCom, improves performance for both Task1 and Task2. In contrast, w.o.CU exhibits a steadier and slower improvement. This behavior is explained by our hybrid-learning strategy. In the early stages of training, the DRE struggles to provide accurate estimates because the encoder in the SU has not yet converged. As training progresses and the SU starts producing more meaningful outputs, the DRE becomes more effective, leading to visible performance gains.

Moreover, the cooperative processing enabled by the CMT-SemCom framework accelerates this convergence by allowing tasks to share semantic information, thereby reducing the number of iterations required to achieve high accuracy.

A similar behavior is observed for the CIFAR dataset, as shown in Fig. \ref{fig:wwo_cu_cifar}. While cooperation still improves performance, the gap between w.CU and w.o.CU is smaller compared to the MNIST case. Nevertheless, the results confirm that even for complex datasets, IoPm-based CMT-SemCom facilitates the task execution performance.

\subsection{Impact of the Channel Constraint}
We investigate the influence of rate-limited wireless channels on the performance of the proposed CMT-SemCom framework by varying the number of the available channel uses ($d$). We note that $d$ is a translation of the rate constraint and a quantitative representation of the limit. The average task execution error rate for both tasks serves as the evaluation metric to capture how constrained bandwidth affects system accuracy.

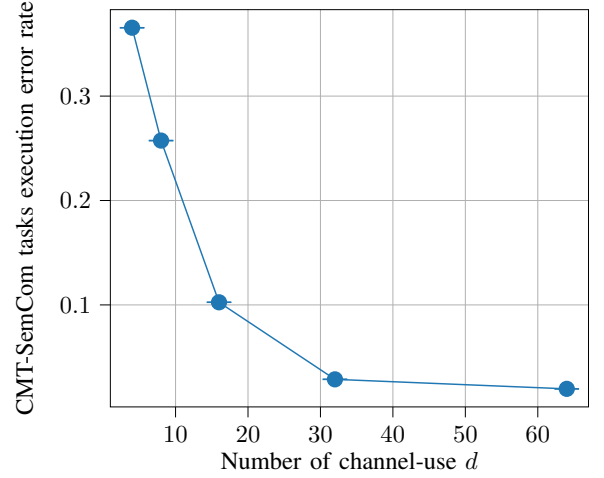
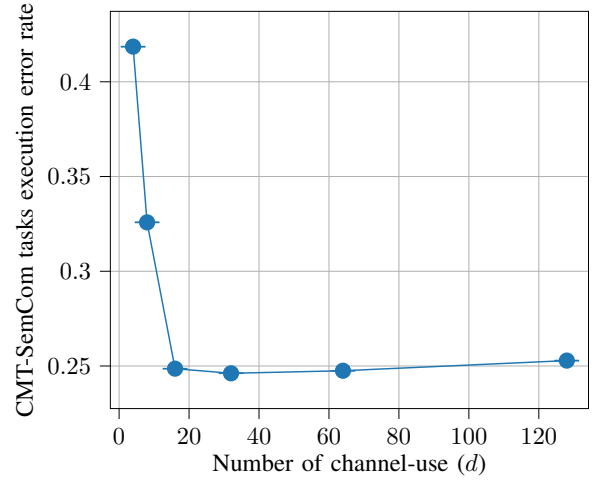
\begin{figure}[t]
    \centering
    \begin{subfigure}{0.95\linewidth}
        \centering
        \resizebox{0.95\textwidth}{!}{
\begin{tikzpicture}

\definecolor{darkgray176}{RGB}{176,176,176}
\definecolor{steelblue31119180}{RGB}{31,119,180}

\begin{axis}[
tick align=outside,
tick pos=left,
x grid style={darkgray176},
xlabel={Number of channel-use $d$},
xmajorgrids,
xmin=1, xmax=67,
xtick style={color=black},
y grid style={darkgray176},
ylabel={CMT-SemCom tasks execution error rate},
ymajorgrids,
ymin=0.00230500400066376, ymax=0.382795003056526,
ytick style={color=black}
]
\path [draw=steelblue31119180, semithick]
(axis cs:64,0.0196000039577484)
--(axis cs:64,0.0196000039577484);

\path [draw=steelblue31119180, semithick]
(axis cs:32,0.0287000238895416)
--(axis cs:32,0.0287000238895416);

\path [draw=steelblue31119180, semithick]
(axis cs:16,0.102500021457672)
--(axis cs:16,0.102500021457672);

\path [draw=steelblue31119180, semithick]
(axis cs:8,0.257400006055832)
--(axis cs:8,0.257400006055832);

\path [draw=steelblue31119180, semithick]
(axis cs:4,0.365500003099442)
--(axis cs:4,0.365500003099442);

\addplot [semithick, steelblue31119180, mark=-, mark size=5, mark options={solid}, only marks]
table {%
64 0.0196000039577484
32 0.0287000238895416
16 0.102500021457672
8 0.257400006055832
4 0.365500003099442
};
\addplot [semithick, steelblue31119180, mark=-, mark size=5, mark options={solid}, only marks]
table {%
64 0.0196000039577484
32 0.0287000238895416
16 0.102500021457672
8 0.257400006055832
4 0.365500003099442
};
\addplot [semithick, steelblue31119180, mark=*, mark size=3, mark options={solid}]
table {%
64 0.0196000039577484
32 0.0287000238895416
16 0.102500021457672
8 0.257400006055832
4 0.365500003099442
};
\end{axis}

\end{tikzpicture}}
        \caption{Performance on MNIST dataset.}
        \label{fig:rate_limit_mnist}
    \end{subfigure}  
    \vspace{0.4em}
    \begin{subfigure}{0.95\linewidth}
        \centering
        \resizebox{0.95\textwidth}{!}{
\begin{tikzpicture}

\definecolor{darkgray176}{RGB}{176,176,176}
\definecolor{steelblue31119180}{RGB}{31,119,180}

\begin{axis}[
tick align=outside,
tick pos=left,
x grid style={darkgray176},
xlabel={Number of channel-use ($d$)},
xmajorgrids,
xmin=-2.5, xmax=134.2,
xtick style={color=black},
y grid style={darkgray176},
ylabel={CMT-SemCom tasks execution error rate},
ymajorgrids,
ymin=0.227529987096786, ymax=0.437169984579086,
ytick style={color=black}
]
\path [draw=steelblue31119180, semithick]
(axis cs:128,0.252849996089935)
--(axis cs:128,0.252849996089935);

\path [draw=steelblue31119180, semithick]
(axis cs:64,0.247500002384186)
--(axis cs:64,0.247500002384186);

\path [draw=steelblue31119180, semithick]
(axis cs:32,0.246149986982346)
--(axis cs:32,0.246149986982346);

\path [draw=steelblue31119180, semithick]
(axis cs:16,0.248600006103516)
--(axis cs:16,0.248600006103516);

\path [draw=steelblue31119180, semithick]
(axis cs:8,0.325800001621246)
--(axis cs:8,0.325800001621246);

\path [draw=steelblue31119180, semithick]
(axis cs:4,0.418549984693527)
--(axis cs:4,0.418549984693527);

\addplot [semithick, steelblue31119180, mark=-, mark size=5, mark options={solid}, only marks]
table {%
128 0.252849996089935
64 0.247500002384186
32 0.246149986982346
16 0.248600006103516
8 0.325800001621246
4 0.418549984693527
};
\addplot [semithick, steelblue31119180, mark=-, mark size=5, mark options={solid}, only marks]
table {%
128 0.252849996089935
64 0.247500002384186
32 0.246149986982346
16 0.248600006103516
8 0.325800001621246
4 0.418549984693527
};
\addplot [semithick, steelblue31119180, mark=*, mark size=3, mark options={solid}]
table {%
128 0.252849996089935
64 0.247500002384186
32 0.246149986982346
16 0.248600006103516
8 0.325800001621246
4 0.418549984693527
};
\end{axis}

\end{tikzpicture}}
        \caption{Performance on CIFAR dataset.}
        \label{fig:rate_limit_cifar}
    \end{subfigure}
    \caption{Impact of the rate limitation of the wireless channel on the proposed CMT-SemCom.}
    \label{fig:placeholder}
\end{figure}

Fig. \ref{fig:rate_limit_mnist} and Fig. \ref{fig:rate_limit_cifar} illustrate this effect for the MNIST and CIFAR datasets, respectively.

For the MNIST dataset, a clear performance degradation is observed as the channel becomes more constrained. The task execution error increases from approximately 2\% at 64 channel uses to over 36\% at 4 channel uses. This behavior highlights the sensitivity of the system's performance to channel limitations. 

For the CIFAR dataset, the behavior differs. While performance degrades at extreme channel limitations (e.g., 4 and 8 channel uses), the error rate remains relatively stable across a broad range. Moreover, the steeper drop in performance below the 16 channel uses for CIFAR dataset indicates a threshold effect. Once the encoded representation faces a specific limit, the loss of information becomes significant, and a sharp drop in task execution quality takes place. The impact is less dramatic for the MNIST.

\subsection{Hybrid-Learning IoPm Vs. EP}
\begin{figure*}[t!]
    \centering
    \begin{subfigure}{0.50\textwidth}
        \centering
        \resizebox{0.85\linewidth}{!}{\input{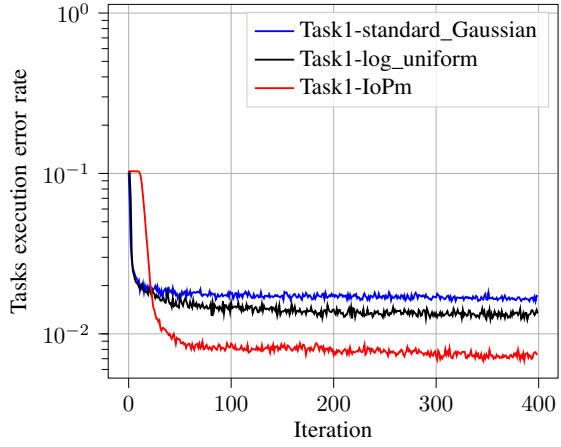}}
        \caption{Comparison for Task1 on MNIST dataset.}
        \label{fig:iopm_ep_mnist_task1}
    \end{subfigure}%
    \hfill
    \begin{subfigure}{0.50\textwidth}
        \centering
        \resizebox{0.85\linewidth}{!}{\input{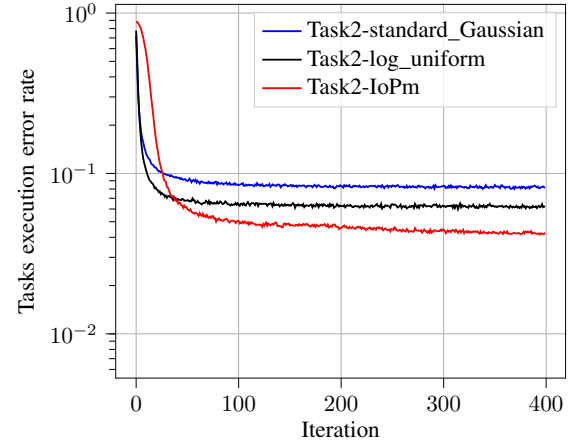}}
        \caption{Comparison for Task2 on MNIST dataset.}
        \label{fig:iopm_ep_mnist_task2}
    \end{subfigure}
    \begin{subfigure}{0.50\textwidth}
        \centering
        \resizebox{0.85\linewidth}{!}{\input{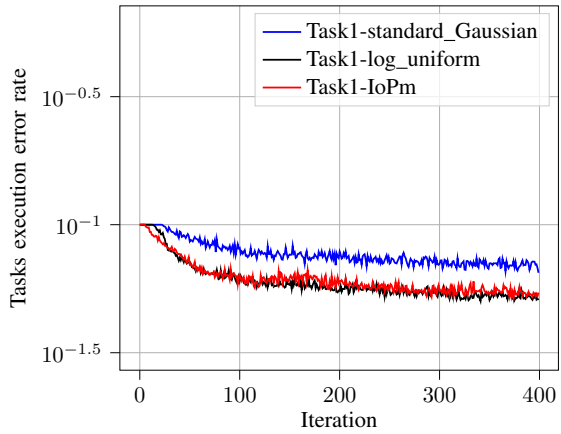}}
        \caption{Comparison for Task1 on CIFAR-10 dataset.}
        \label{fig:iopm_ep_cifar_task1}
    \end{subfigure}%
    \hfill
    \begin{subfigure}{0.50\textwidth}
        \centering
        \resizebox{0.85\linewidth}{!}{\input{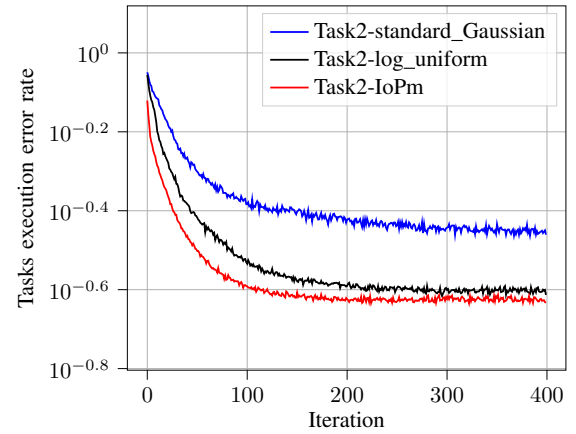}}
        \caption{Comparison for Task2 on CIFAR-10 dataset.}
        \label{fig:iopm_ep_cifar_task2}
    \end{subfigure}
    
    \caption{Comparison of the proposed hybrid-learning IoPm with the EP method on MNIST and CIFAR-10 datasets.}
    \label{fig:iopm_ep}
\end{figure*}

Finally, we compare our proposed hybrid-learning based IoPm approach with the two widely used fixed priors in EP method, which are the standard Gaussian and the log-uniform prior. As shown in Figs. \ref{fig:iopm_ep_mnist_task1} and \ref{fig:iopm_ep_mnist_task2}, our hybrid-learning-based IoPm consistently outperforms the EP method for both tasks in the MNIST dataset, resulting in improved execution accuracy for both tasks. We observe in both figures that the performance gap between the IoPm and the fixed standard Gaussian prior is larger than the gap between the IoPm and the fixed log-uniform. This reflects the larger error of the model assumption with the Gaussian distribution for the encoded feature. This demonstrates the advantage of using a learned, data-driven prior over a fixed distribution.

For the complex dataset, CIFAR, we observe the same performance gain for the IoPm in comparison with the standard Gaussian prior for both tasks. This is shown in Figs. \ref{fig:iopm_ep_cifar_task1} and \ref{fig:iopm_ep_cifar_task2}. In addition, we note that the gap between the IoPm and the standard Gaussian grows when the task becomes more complex (as for Task2, Fig. \ref{fig:iopm_ep_cifar_task2}) and as a result its corresponding encoded feature is more complicated. Further, comparing the IoPm with the log-uniform prior, Fig. \ref{fig:iopm_ep_cifar_task2} shows that the proposed method maintains its better performance for Task2, while for the simpler one (as Task1, Fig. \ref{fig:iopm_ep_cifar_task1})the performance difference is less pronounced. This is due to the fact that the log-uniform assumption fits well in modeling the encoded feature distribution. 

Overall, we observe that the proposed hybrid-learning IoPm reaches clear performance gains in comparison to the EP method. This performance gain is more considerable compared to the widely used explicit standard Gaussian prior independent of the dataset. The proposed method also mitigates the model assumption error for the latent prior in related domains, e.g., variational autoencoders, information bottleneck-based task-oriented communication, etc.

We also note that the extra computational cost introduced by the IoPm is incurred only during offline training. Therefore, the proposed method does not increase complexity in inference time as the DRE is used only during the training for a more accurate approximation of the KL divergence term. A detailed discussion on the comparison of computational complexity between the IoPm and the EP method is provided in Appendix \ref{appendix:DRE_computational}.

\section{Conclusion} \label{section.Conclusion}

In conclusion, we advanced the CMT-SemCom framework by addressing practical constraints and extending its applicability to rate-limited wireless channels. We employed a separation-based design for the split semantic encoder to have a clear delineation of responsibilities between the CU and SUs, facilitating a more structured formulation of the communication process. Further, we tackled the regularization challenge within the joint semantic and channel coding process by employing the implicit optimal prior method (IoPm) to enhance the system's performance. We proposed a hybrid combination of DNN and kernelized-parametric ML methods to improve the approximation of the constrained problem. Through simulations on diverse datasets, we demonstrated the effectiveness of the proposed framework in achieving reliable multi-task communication under rate constraints compared to explicit prior (EP) method. Our comparisons with EP include the two widely used fixed priors in the related literature: the standard Gaussian prior and the log-uniform. These two explicit priors are widely used in the literature to model the encoded/latent feature prior. Additionally, this work brings up further research questions, such as exploring alternative parametric methods for the DRE, like the ratio matching method instead of the kernelized LR, or examining dimensionality reduction techniques to better exploit the current DRE’s capabilities and improve performance in complex settings where higher dimensionality is required, i.e., more complex tasks.  dynamic adaptation of semantic encoding structures to varying network requirements.

\appendices

\section{Differentiability of the CU Loss Function} \label{Appendix_CU-reparameterization}
Here we first show the differentiability of (\ref{eq:CU-objective-final}) w.r.t $\boldsymbol{\theta}$ and then details on the reparameterization trick are provided.
\subsection{Derivative w.r.t the CU Encoder Parameters} \label{appendix:derivative}
The differentiability of the CU loss function w.r.t $\boldsymbol{\Xi}$ is clear since it is explicitly stated in (\ref{eq:CU-objective-final}), however how $\boldsymbol{\theta}$ is updated through the backpropagation is not explicitly visible. Thus, below we show how (\ref{eq:CU-objective-final}) is differentiable w.r.t $\boldsymbol{\theta}$.
\begin{equation*}
    \mathcal{L}^{\text{\tiny CU}}(\boldsymbol{\theta}, \boldsymbol{\Xi}) \approx \mathbb{E}_{p(\mathbf{S},\mathbf{z})} \left[\ \mathbb{E}_{p^{\text{\tiny CU}}_{\boldsymbol{\theta}}(\mathbf{c}|\mathbf{S})} [\ f(\mathbf{z})\, ]\,\right]
\end{equation*}
Where $\mathbf{z}=g(\mathbf{c},\boldsymbol{\Xi})$, and $\mathbf{c}=h(\mathbf{S},\boldsymbol{\theta},\boldsymbol{\epsilon})$. Thus, (\ref{eq:CU-objective-final}) can be expressed as:
\begin{equation*}
    \mathcal{L}^{\text{\tiny CU}}(\boldsymbol{\theta}, \boldsymbol{\Xi}) \approx \mathbb{E}_{p(\mathbf{S},\mathbf{z})} \left[\ \mathbb{E}_{p^{\text{\tiny CU}}_{\boldsymbol{\theta}}(\mathbf{c}|\mathbf{S})} [\ f(g(h(\mathbf{S},\boldsymbol{\theta},\boldsymbol{\epsilon}),\boldsymbol{\Xi}))\, ]\,\right]
\end{equation*}
Consequently:
\begin{equation*}
\begin{aligned}
    &\mathcal{L}^{\text{\tiny CU}}(\boldsymbol{\theta}, \boldsymbol{\Xi}) \approx f(g(h(\mathbf{S},\boldsymbol{\theta},\boldsymbol{\epsilon}),\boldsymbol{\Xi})) \\[0.8em]
    &\frac{\partial\mathcal{L}^{\text{\tiny CU}}(\boldsymbol{\theta}, \boldsymbol{\Xi})}{\partial \boldsymbol{\theta}} = \frac{\partial f}{\partial g} \cdot \frac{\partial g}{\partial h} \cdot \frac{\partial h}{\partial \boldsymbol{\theta}}
\end{aligned}
\end{equation*}
\subsection{Reparameterization Trick} \label{appendix:reparameterization}
We assume that $p^{\text{\tiny CU}}_{\boldsymbol{\theta}}(\mathbf{c}|\mathbf{S})=\mathcal{N}(\mathbf{c}(\mathbf{S};\boldsymbol{\theta}), \sigma^2 \mathbf{I})$, where $\mathbf{c}(\mathbf{S};\boldsymbol{\theta})$ states the deterministic function which maps $\mathbf{S}$ to $\mathbf{c}$ parameterized by $\boldsymbol{\theta}$. It is obvious that $\mathbf{c} \sim p^{\text{\tiny CU}}_{\boldsymbol{\theta}}(\mathbf{c}|\mathbf{S})$ and then in backpropagation when the update w.r.t $\boldsymbol{\theta}$ wants to be executed there will be a problem by:
\begin{equation*}
    \nabla_{\boldsymbol{\theta}} \mathbb{E}_{p^{\text{\tiny CU}}_{\boldsymbol{\theta}}(\mathbf{c}|\mathbf{S})} [\ f(g(c(\mathbf{S};\boldsymbol{\theta}),\boldsymbol{\Xi}))\, ]\
\end{equation*}
Therefore, we introduce a new variable $\boldsymbol{\epsilon}$ as $\mathbf{c}_{i,l} = \mathbf{c}_i + \boldsymbol{\epsilon}_{i,l}$, where we keep $\mathbf{c}_i$ a deterministic variable and $\boldsymbol{\epsilon}_{i,l}$ a sample drawn from $\mathcal{N}(\mathbf{0}, \sigma^2 \mathbf{I})$ distribution. Doing so, the expectation would be w.r.t $p(\boldsymbol{\epsilon})$ as follows, and the differentiability w.r.t $\boldsymbol{\theta}$ will be possible.
\begin{equation*}
\begin{aligned}
    & \nabla_{\boldsymbol{\theta}} \mathbb{E}_{p^{\text{\tiny CU}}_{\boldsymbol{\theta}}(\mathbf{c}|\mathbf{S})} [\ f(g(c(\mathbf{s};\boldsymbol{\theta}),\boldsymbol{\Xi}))\,]\ \\[0.6em]
    &= \nabla_{\boldsymbol{\theta}} \mathbb{E}_{p(\boldsymbol{\epsilon})} [\ f(g(c(\boldsymbol{\epsilon},\mathbf{S};\boldsymbol{\theta}),\boldsymbol{\Xi}))\, ]\ \\[0.8em]
    &=\mathbb{E}_{p(\boldsymbol{\epsilon})} [\ \nabla_{\boldsymbol{\theta}} f(g(c(\boldsymbol{\epsilon},\mathbf{S};\boldsymbol{\theta}),\boldsymbol{\Xi}))\, ]\ \\[0.8em]
    &\simeq \nabla_{\boldsymbol{\theta}} f(g(c(\boldsymbol{\epsilon},\mathbf{S};\boldsymbol{\theta}),\boldsymbol{\Xi}))
\end{aligned}
\end{equation*}

\section{The Approximated SUs' Objective Function} \label{Appendix:su_loss}

\begin{equation*}
    \begin{aligned}
        \mathcal{L}^{\text{\tiny SU$_n$}}(\boldsymbol{\phi}_n) &=  I(\bm{\hat{x}}_n;z_n) - \lambda\,I(\bm{x}_n;\bm{c}) \\[0.6em]
        &= \iint p(\bm{\hat{x}}_n, z_n)\,\log \frac{p(z_n|\bm{\hat{x}}_n)}{p(z_n)}\,dz_n \, d\bm{\hat{x}}_n \\[0.6em]
        & \quad - \lambda \bigg(\ \iint p(\bm{x}_n, \bm{c})\,\log \frac{p^{\text{\tiny SU}}_{\boldsymbol{\phi}_n}(\bm{x}_n|\bm{c})}{p(\bm{x}_n)}\,d\bm{x}_n \, d\bm{c} \bigg)\ \\[0.6em]
        &\hspace{-0.6em} = \left [\ \iint p(\bm{\hat{x}}_n, z_n)\,\log \,p(z_n|\bm{\hat{x}}_n)\,dz_n \, d\bm{\hat{x}}_n + H(z_n)\right]\ \\[0.6em]
        & \quad - \lambda \bigg(\ \iint p(\bm{x}_n, \bm{c})\,\log \frac{p^{\text{\tiny SU}}_{\boldsymbol{\phi}_n}(\bm{x}_n|\bm{c})}{p(\bm{x}_n)}\,d\bm{x}_n \, d\bm{c} \bigg)\ \\[0.6em] 
    \end{aligned}
\end{equation*}

Further, we omit the constant entropy term, $H(z_n)$ and exploit the underlying Markov chain structure in (\ref{eq:system_probability}).
\begin{equation*}
    \begin{aligned}
        \mathcal{L}^{\text{\tiny SU$_n$}}(\boldsymbol{\phi}_n) &\approx \iiiint p_{\boldsymbol{\theta}}(z_n,\bm{c})\,p^{\text{\tiny SU}}_{\boldsymbol{\phi}_n}(\bm{x}_n|\mathbf{c}) \\[0.6em] 
        & \qquad \quad \quad p(\bm{\hat{x}}_n|\bm{x}_n) \log p(z_n|\bm{\hat{x}}_n)\, dz_n\, d\bm{\hat{x}}_n\, d\bm{x}_n\, d\bm{c} \\[0.6em]
        & \quad - \lambda \bigg(\ \iiint p_{\boldsymbol{\theta}}(z_n,\bm{c})\\[0.6em]
        &\qquad \qquad p^{\text{\tiny SU}}_{\boldsymbol{\phi}_n}(\bm{x}_n| \bm{c})\,\log \frac{p^{\text{\tiny SU}}_{\boldsymbol{\phi}_n}(\bm{x}_n|\bm{c})}{p(\bm{x}_n)}\,dz_n\,d\bm{x}_n \, d\bm{c} \bigg)\ \\[0.6em]
        & \approx \mathbb{E}_{p_{\boldsymbol{\theta}}(z_n,\bm{c})}\left[\, \iint p^{\text{\tiny SU}}_{\boldsymbol{\phi}_n}(\bm{x}_n|\mathbf{c}) p(\bm{\hat{x}}_n|\bm{x}_n) \right. \\[0.6em]
        & \qquad \qquad \qquad \left. \log p(z_n|\bm{\hat{x}}_n)\,d\bm{\hat{x}}_n\, d\bm{x}_n \right. \\[0.6em] 
        & \qquad \left. - \lambda \int p^{\text{\tiny SU}}_{\boldsymbol{\phi}_n}(\bm{x}_n| \bm{c})\,\log \frac{p^{\text{\tiny SU}}_{\boldsymbol{\phi}_n}(\bm{x}_n|\bm{c})}{p(\bm{x}_n)}\,d\bm{x}_n \right]\
    \end{aligned}
\end{equation*}

Next, we adopt the definition of the KL \cite{cover1999elements},
\begin{equation*}
    KL(f||g) = \int f \log \frac{f}{g},
\end{equation*}
and the approximated loss function is expressed as:
\begin{equation*}
    \begin{aligned}
        &\mathcal{L}^{\text{\tiny SU$_n$}}(\boldsymbol{\phi}_n) \\[0.6em] &\approx \mathbb{E}_{p_{\boldsymbol{\theta}}(z_n,\bm{c})} \left[\ \mathbb{E}_{p^{\text{\tiny SU}}_{\boldsymbol{\phi}_n}(\bm{x}_n|\bm{c})} \left[\,\mathbb{E}_{p(\bm{\hat{x}}_n|\bm{x}_n)} [\ \log p(z_n|\bm{\hat{x}}_n)\, ]\ \right]\ \right.\\[0.6em] 
        & \qquad \qquad \qquad \qquad - \left. \lambda\, KL(p^{\text{\tiny SU}}_{\boldsymbol{\phi}_n}(\bm{x}_n|\bm{c})\parallel p(\bm{x}_n))\right]\,.
    \end{aligned}
\end{equation*}

It is important to note that $p_{\boldsymbol{\theta}}(z_n,\bm{c})$ is readily available at this stage, owing to the pre-trained CU. In essence, we construct our Markov chain by treating $p_{\boldsymbol{\theta}}(z_n,\bm{c})$ as a new, derived source distribution.

\section{SU Objective Approximation Error Analysis} \label{appendix:SU_approx_error}

In (\ref{eq:SU-objective-final}), $p^{\text{\tiny SU}}(\bm{x}_n|\bm{c})$, called SU encoder, is modeled in terms of a DNN parameterized by $\boldsymbol{\phi}_n$. Thus, the encoder is treated as part of model selection, and its error is about \emph{model mismatch}. The approximation error it brings to the loss function $\mathcal{L}^{\text{\tiny SU$_n$}}(\boldsymbol{\phi}_n, \boldsymbol{\psi}_n)$, is reflected through the KL divergence term $KL\left(p^{\text{\tiny SU}}_{\boldsymbol{\phi}_n}(\bm{x}_n|\bm{c})\parallel p(\bm{x}_n)\right)$. The variational approximation error comes from the decoder, where the true posterior, $p(z_n|\hat{\bm{x}}_n)$ is approximated by $q_{\boldsymbol{\psi}_n}(z_n|\hat{\bm{x}}_n)$. The mismatch between the true and approximated posterior indicated the error in the variational approximation, and therefore, the loss will have another approximation error through the log-likelihood function (LLF).

Thus, the approximation error for the objective consists of two terms:
\begin{itemize}
    \item LLF variational approximation error
    \item Model mismatch error
\end{itemize}

We begin with the LLF variational approximation error. We use a variant of Pinsker's inequality \cite{pinsker1964} which relates variational divergence to KL divergence as introduced in \cite{vaeapproxerror}:
    \begin{equation*}
    \sup_{A \subseteq \mathcal{Z}} |P(A)-Q(A)|^2 \leq \frac{1}{2} \, KL(P||Q),
    \end{equation*}
substituting the probabilities $P$ and $Q$ with the decoder's true and approximated ones:
    \begin{equation*}
         |p(z_n|\hat{\bm{x}}_n)-q_{\boldsymbol{\psi}_n}(z_n|\hat{\bm{x}}_n)|^2 \leq \frac{1}{2} \, KL(p(z_n|\hat{\bm{x}}_n)||q_{\boldsymbol{\psi}_n}(z_n|\hat{\bm{x}}_n)),
    \end{equation*}
and, similar to the definition assumed for the evidence lower bound, for instance, in variational autoencoders \cite{kingma2019introduction,vaeapproxerror}, we define our approximation quality as $\epsilon$-tight for some $\epsilon > 0$ if
\begin{equation*}
    \mathbb{E}_{p^{\text{\tiny SU}}_{\boldsymbol{\phi}_n}(\hat{\bm{x}}_n|\bm{c})} \left[\, KL\left(p(z_n|\hat{\bm{x}}_n)||q_{\boldsymbol{\psi}_n}(z_n|\hat{\bm{x}}_n)\right) \right]\, \leq \epsilon,    
\end{equation*}
further, we assume $p(z_n|\hat{\bm{x}}_n) \geq \alpha$ and $q_{\boldsymbol{\psi}_n}(z_n|\hat{\bm{x}}_n) \geq \alpha$.

Thus, taking the expectation from both sides of the Pinsker's inequality, we bound the variational approximation error for the LLF term as:
    \begin{equation*}
       \mathbb{E}_{p^{\text{\tiny SU}}_{\boldsymbol{\phi}_n}(\hat{\bm{x}}_n|\bm{c})} \left[\ |\log p(z_n|\hat{\bm{x}}_n) - \log q_{\boldsymbol{\psi}_n}(z_n|\hat{\bm{x}}_n)|^2 \right]\ \leq \frac{\epsilon}{2\alpha^2} + o(\epsilon).
    \end{equation*}

The detailed proof of getting this bound has been provided in \cite{vaeapproxerror}. Next, applying the Cauchy-Schwarz inequality ($|\mathbb{E}[\ XY ]\,|^2 \leq \mathbb{E}[\ X^2]\ \mathbb{E}[\ Y^2]\ $) to the error bound, we get LLF's error to our loss as follows:
\begin{equation*}
    \begin{aligned}
       &|\mathbb{E}_{p^{\text{\tiny SU}}_{\boldsymbol{\phi}_n}(\hat{\bm{x}}_n|\bm{c})} \left[\ \log p(z_n|\hat{\bm{x}}_n) - \log q_{\boldsymbol{\psi}_n}(z_n|\hat{\bm{x}}_n) \right]\,| \leq \\[0.6em] & \sqrt{\mathbb{E}_{p^{\text{\tiny SU}}_{\boldsymbol{\phi}_n}(\hat{\bm{x}}_n|\bm{c})} \left[\ |\log p(z_n|\hat{\bm{x}}_n) - \log q_{\boldsymbol{\psi}_n}(z_n|\hat{\bm{x}}_n)|^2 \right]\,} \leq \\[0.6em]  
        & \sqrt{\frac{\epsilon}{2\alpha^2} + o(\epsilon)} \leq \sqrt{\frac{\epsilon}{2\alpha^2}}
    \end{aligned}
\end{equation*}

Next, we move to the model mismatch. We once more use the assumption of $\epsilon^\prime$-tightness for $\epsilon^\prime > 0$ in the model selection. It has been shown in \cite{dai2019diagnosing} that for the continuous case with a non-linear encoder/decoder (general neural networks), this $\epsilon^\prime$-tightness is satisfied. Therefore, we bound the error for our model approximation by:
    \begin{equation*}
        \mathbb{E}_{p_{\boldsymbol{\theta}}(\bm{c})} \left[\,KL\left(p^{\text{\tiny SU}}_{\boldsymbol{\phi}_n}(\bm{x}_n|\bm{c})\parallel p(\bm{x}_n|\bm{c})\right) \right]\, \leq \epsilon^\prime
    \end{equation*}
Finally, considering $\mathcal{L}^{\text{\tiny SU$_n$}}$ as the true objective function, the approximation error for the objective in (\ref{eq:SU-objective-final}) is expressed by the following upper-bound:
    \begin{equation*}
        |\mathcal{L}^{\text{\tiny SU$_n$}} - \mathcal{L}^{\text{\tiny SU$_n$}}(\boldsymbol{\phi}_n, \boldsymbol{\psi}_n)| \leq \sqrt{\frac{\epsilon}{2\alpha^2}} \, + \,\epsilon^\prime
    \end{equation*}
    
\section{Backpropagation of the SU Loss Function} \label{appendix:SU_loss_gradients}
Here we show why we ignore the $\hat{r}(\mathbf{x}_n)$ in the backpropagation of the approximated objective function of the n-th SU in \ref{eq:SU-objective-with-DRE-applied}. We use the SGD over mini-batches to train our SU encoder, and the update procedure looks:
\begin{equation*}
    \boldsymbol{\phi}^{\tau + 1}_{n} \leftarrow \boldsymbol{\phi}^{\tau}_{n} - \frac{\alpha}{M_\tau} \sum_{i\in M_\tau} \nabla_{\boldsymbol{\phi}_n}\mathcal{L}^{\text{\tiny SU$_n$}}_{i}(\boldsymbol{\phi}^{\tau}_{n}, \boldsymbol{\psi}^{\tau}_{n})
\end{equation*}
Thus, when optimized DRE is employed in the objective function, the gradient term becomes zero, and that is why we ignore the involvement of $\hat{r}(\mathbf{x}_n)$ in the optimization of the SU encoder. It is obvious that for other non-linear optimization techniques, such as Adam, the ignorance of the DRE in the updating step of the SUs is not possible.
\begin{equation*}
    \begin{aligned}
        \mathbb{E}_{p_{\boldsymbol{\phi}_n}(\mathbf{x}_n)} \left[\,\nabla_{\boldsymbol{\phi}_n}\log p_{\boldsymbol{\phi}_n}(\mathbf{x}_n) \right]\ &= \int p_{\boldsymbol{\phi}_n}(\mathbf{x}_n) \frac{\nabla_{\boldsymbol{\phi}_n}p_{\boldsymbol{\phi}_n}(\mathbf{x}_n)}{p_{\boldsymbol{\phi}_n}(\mathbf{x}_n)}\,d\mathbf{x}_n\\[0.6em]
        &= \nabla_{\boldsymbol{\phi}_n}\int p_{\boldsymbol{\phi}_n}(\mathbf{x}_n)\,d\mathbf{x}_n = 0
    \end{aligned}
\end{equation*}

\section{IoPm Computational Complexity} \label{appendix:DRE_computational}
On training computational complexity, let $\bm{m}$ be the number of the DRE training samples, and $\bm{d}$ be the dimensionality of the SU encoder output. The computational complexity of our Gaussian kernel, represented in eq. (\ref{eq:kernel}), will be $\mathcal{O}(m^2d)$, which is the operations for computing the kernel Gram matrix, and therefore, $\mathcal{O}(m^2)$ for storing the matrix.

In addition, the LR classifier is trained for $\bm{T}$ iterations of gradient descent over the samples, and the total computational complexity of the training using the transformed kernel will be $\mathcal{O}(Tm^2)$. In contrast, the EP methods (fixed standard Gaussian and log-uniform priors), with which we compared our proposed IoPm, do not require the DRE and therefore have lower training complexity and training time. However, as mentioned earlier, since the DRE is present only during offline training, the training complexity is manageable due to the possibility of using powerful resources.

\bibliographystyle{IEEEtran}
\bibliography{IEEEabrv,References.bib}

\begin{thebibliography}{10}
\providecommand{\url}[1]{#1}
\csname url@samestyle\endcsname
\providecommand{\newblock}{\relax}
\providecommand{\bibinfo}[2]{#2}
\providecommand{\BIBentrySTDinterwordspacing}{\spaceskip=0pt\relax}
\providecommand{\BIBentryALTinterwordstretchfactor}{4}
\providecommand{\BIBentryALTinterwordspacing}{\spaceskip=\fontdimen2\font plus
\BIBentryALTinterwordstretchfactor\fontdimen3\font minus \fontdimen4\font\relax}
\providecommand{\BIBforeignlanguage}[2]{{%
\expandafter\ifx\csname l@#1\endcsname\relax
\typeout{** WARNING: IEEEtran.bst: No hyphenation pattern has been}%
\typeout{** loaded for the language `#1'. Using the pattern for}%
\typeout{** the default language instead.}%
\else
\language=\csname l@#1\endcsname
\fi
#2}}
\providecommand{\BIBdecl}{\relax}
\BIBdecl

\bibitem{10654356}
A.~Halimi~Razlighi, C.~Bockelmann, and A.~Dekorsy, ``Semantic communication for cooperative multi-task processing over wireless networks,'' \emph{IEEE Wireless Communications Letters}, vol.~13, no.~10, pp. 2867--2871, 2024.

\bibitem{Gunduz2022}
D.~Gündüz, Z.~Qin, I.~E. Aguerri, H.~S. Dhillon, Z.~Yang, A.~Yener, K.~K. Wong, and C.-B. Chae, ``Beyond transmitting bits: Context, semantics, and task-oriented communications,'' \emph{IEEE Journal on Selected Areas in Communications}, vol.~41, no.~1, pp. 5--41, 2023.

\bibitem{Luo2022}
X.~Luo, H.~H. Chen, and Q.~Guo, ``Semantic communications: Overview, open issues, and future research directions,'' \emph{IEEE Wireless Communications}, vol.~29, pp. 210--219, 2 2022.

\bibitem{CALVANESESTRINATI2021107930}
\BIBentryALTinterwordspacing
E.~{Calvanese Strinati} and S.~Barbarossa, ``6{G} networks: Beyond shannon towards semantic and goal-oriented communications,'' \emph{Computer Networks}, vol. 190, p. 107930, 2021. [Online]. Available: \url{https://www.sciencedirect.com/science/article/pii/S1389128621000773}
\BIBentrySTDinterwordspacing

\bibitem{qin2021semantic}
Z.~Qin, X.~Tao, J.~Lu, W.~Tong, and G.~Y. Li, ``Semantic communications: Principles and challenges,'' \emph{arXiv preprint arXiv:2201.01389}, 2021.

\bibitem{WenTong}
W.~Tong and G.~Y. Li, ``Nine challenges in artificial intelligence and wireless communications for {6G},'' \emph{IEEE Wireless Communications}, vol.~29, no.~4, pp. 140--145, 2022.

\bibitem{Shannon1948}
C.~E. Shannon, ``A mathematical theory of communication,'' 1948.

\bibitem{Sana2022}
M.~Sana and E.~C. Strinati, ``Learning semantics: An opportunity for effective 6{G} communications.''\hskip 1em plus 0.5em minus 0.4em\relax Institute of Electrical and Electronics Engineers Inc., 2022, pp. 631--636.

\bibitem{Wheeler2023}
D.~Wheeler and B.~Natarajan, ``Engineering semantic communication: A survey,'' \emph{IEEE Access}, vol.~11, pp. 13\,965--13\,995, 2023.

\bibitem{weaver1953recent}
W.~Weaver, ``Recent contributions to the mathematical theory of communication,'' \emph{ETC: a review of general semantics}, pp. 261--281, 1953.

\bibitem{BarHillelCarnap}
\BIBentryALTinterwordspacing
Y.~Bar-Hillel and R.~Carnap, ``Semantic information,'' \emph{The British Journal for the Philosophy of Science}, vol.~4, no.~14, pp. 147--157, 1953. [Online]. Available: \url{http://www.jstor.org/stable/685989}
\BIBentrySTDinterwordspacing

\bibitem{Wang2017}
Q.~Wang, Z.~Mao, B.~Wang, and L.~Guo, ``Knowledge graph embedding: A survey of approaches and applications,'' \emph{IEEE Transactions on Knowledge and Data Engineering}, vol.~29, pp. 2724--2743, 12 2017.

\bibitem{Zhou2022}
F.~Zhou, Y.~Li, X.~Zhang, Q.~Wu, X.~Lei, and R.~Q. Hu, ``Cognitive semantic communication systems driven by knowledge graph,'' in \emph{ICC 2022 - IEEE International Conference on Communications}, 2022, pp. 4860--4865.

\bibitem{Xie2021}
H.~Xie, Z.~Qin, G.~Y. Li, and B.~H. Juang, ``Deep learning enabled semantic communication systems,'' \emph{IEEE Transactions on Signal Processing}, vol.~69, pp. 2663--2675, 2021.

\bibitem{Xie2021-2}
H.~Xie and Z.~Qin, ``A lite distributed semantic communication system for internet of things,'' \emph{IEEE Journal on Selected Areas in Communications}, vol.~39, pp. 142--153, 1 2021.

\bibitem{10599525MBM}
L.~Qiao, M.~B. Mashhadi, Z.~Gao, C.~H. Foh, P.~Xiao, and M.~Bennis, ``Latency-aware generative semantic communications with pre-trained diffusion models,'' \emph{IEEE Wireless Communications Letters}, vol.~13, no.~10, pp. 2652--2656, 2024.

\bibitem{xu2024semanticawarepowerallocationgenerative}
\BIBentryALTinterwordspacing
C.~Xu, M.~B. Mashhadi, Y.~Ma, and R.~Tafazolli, ``Semantic-aware power allocation for generative semantic communications with foundation models,'' 2024. [Online]. Available: \url{https://arxiv.org/abs/2407.03050}
\BIBentrySTDinterwordspacing

\bibitem{Yan2022}
L.~Yan, Z.~Qin, R.~Zhang, Y.~Li, and G.~Y. Li, ``Resource allocation for text semantic communications,'' \emph{IEEE Wireless Communications Letters}, vol.~11, pp. 1394--1398, 7 2022.

\bibitem{Wang2022}
\BIBentryALTinterwordspacing
Y.~Wang, S.~Member, M.~Chen, T.~Luo, S.~Member, W.~Saad, D.~Niyato, H.~V. Poor, L.~Fellow, S.~Cui, and P.~Cheng, ``Performance optimization for semantic communications: An attention-based reinforcement learning approach and also with the,'' \emph{IEEE JOURNAL ON SELECTED AREAS IN COMMUNICATIONS}, vol.~40, 2022. [Online]. Available: \url{https://www.ieee.org/publications/rights/index.html}
\BIBentrySTDinterwordspacing

\bibitem{Tong2021}
H.~Tong, Z.~Yang, S.~Wang, Y.~Hu, W.~Saad, and C.~Yin, ``Federated learning based audio semantic communication over wireless networks.''\hskip 1em plus 0.5em minus 0.4em\relax Institute of Electrical and Electronics Engineers Inc., 2021.

\bibitem{9685056KG}
Y.~Wang, M.~Chen, W.~Saad, T.~Luo, S.~Cui, and H.~V. Poor, ``Performance optimization for semantic communications: An attention-based learning approach,'' in \emph{2021 IEEE Global Communications Conference (GLOBECOM)}, 2021, pp. 1--6.

\bibitem{JammingSemantic}
Y.~Sun, Z.~Lin, K.~An, D.~Li, C.~Li, Y.~Zhu, D.~Wing Kwan~Ng, N.~Al-Dhahir, and J.~Wang, ``Multi-functional ris-assisted semantic anti-jamming communication and computing in integrated aerial-ground networks,'' \emph{IEEE Journal on Selected Areas in Communications}, vol.~42, no.~12, pp. 3597--3617, 2024.

\bibitem{wu2020understandingMTL}
S.~Wu, H.~R. Zhang, and C.~R{\'e}, ``Understanding and improving information transfer in multi-task learning,'' \emph{arXiv preprint arXiv:2005.00944}, 2020.

\bibitem{Shao2021}
J.~Shao, Y.~Mao, and J.~Zhang, ``Learning task-oriented communication for edge inference: An information bottleneck approach,'' \emph{IEEE Journal on Selected Areas in Communications}, vol.~40, no.~1, pp. 197--211, 2022.

\bibitem{Shao2022}
J.~Shao, Y.~Mao, and J.~Zhang, ``Task-oriented communication for multidevice cooperative edge inference,'' \emph{IEEE Transactions on Wireless Communications}, vol.~22, no.~1, pp. 73--87, 2023.

\bibitem{Beck2023}
\BIBentryALTinterwordspacing
E.~Beck, C.~Bockelmann, and A.~Dekorsy, ``Semantic information recovery in wireless networks,'' \emph{Sensors}, vol.~23, p. 6347, 7 2023. [Online]. Available: \url{https://www.mdpi.com/1424-8220/23/14/6347}
\BIBentrySTDinterwordspacing

\bibitem{Task_SensingComm}
D.~Wen, P.~Liu, G.~Zhu, Y.~Shi, J.~Xu, Y.~C. Eldar, and S.~Cui, ``Task-oriented sensing, computation, and communication integration for multi-device edge ai,'' in \emph{ICC 2023 - IEEE International Conference on Communications}, 2023, pp. 3608--3613.

\bibitem{xie2022task}
H.~Xie, Z.~Qin, X.~Tao, and K.~B. Letaief, ``Task-oriented multi-user semantic communications,'' \emph{IEEE Journal on Selected Areas in Communications}, vol.~40, no.~9, pp. 2584--2597, 2022.

\bibitem{he2022learning}
G.~He, S.~Cui, Y.~Dai, and T.~Jiang, ``Learning task-oriented channel allocation for multi-agent communication,'' \emph{IEEE Transactions on Vehicular Technology}, vol.~71, no.~11, pp. 12\,016--12\,029, 2022.

\bibitem{10013075}
Y.~Sheng, F.~Li, L.~Liang, and S.~Jin, ``A multi-task semantic communication system for natural language processing,'' in \emph{2022 IEEE 96th Vehicular Technology Conference (VTC2022-Fall)}, 2022, pp. 1--5.

\bibitem{10520522}
Y.~E. Sagduyu, T.~Erpek, A.~Yener, and S.~Ulukus, ``Multi - receiver task-oriented communications via multi - task deep learning,'' in \emph{2023 IEEE Future Networks World Forum (FNWF)}, 2023, pp. 1--6.

\bibitem{gong2023scalable}
M.~Gong, S.~Wang, and S.~Bi, ``A scalable multi-device semantic communication system for multi-task execution,'' in \emph{GLOBECOM 2023-2023 IEEE Global Communications Conference}.\hskip 1em plus 0.5em minus 0.4em\relax IEEE, 2023, pp. 2227--2232.

\bibitem{caruana1997multitask}
R.~Caruana, ``Multitask learning,'' \emph{Machine learning}, vol.~28, pp. 41--75, 1997.

\bibitem{halimirazlighi2024}
\BIBentryALTinterwordspacing
A.~Halimi~Razlighi, M.~Tillmann, E.~Beck, C.~Bockelmann, and A.~Dekorsy, ``Cooperative and collaborative multi-task semantic communication for distributed sources,'' 2024. [Online]. Available: \url{https://arxiv.org/abs/2411.02150}
\BIBentrySTDinterwordspacing

\bibitem{10502352TWC}
L.~Yan, Z.~Qin, C.~Li, R.~Zhang, Y.~Li, and X.~Tao, ``Qoe-based semantic-aware resource allocation for multi-task networks,'' \emph{IEEE Transactions on Wireless Communications}, vol.~23, no.~9, pp. 11\,958--11\,971, 2024.

\bibitem{kingma2013auto}
D.~P. Kingma and M.~Welling, ``Auto-encoding variational bayes,'' \emph{arXiv preprint arXiv:1312.6114}, 2013.

\bibitem{koller2009probabilistic}
D.~Koller and N.~Friedman, \emph{Probabilistic graphical models: principles and techniques}.\hskip 1em plus 0.5em minus 0.4em\relax MIT press, 2009.

\bibitem{bishop2006pattern}
C.~M. Bishop, ``Pattern recognition and machine learning,'' \emph{Springer google schola}, vol.~2, pp. 1122--1128, 2006.

\bibitem{alemi2016deep}
A.~A. Alemi, I.~Fischer, J.~V. Dillon, and K.~Murphy, ``Deep variational information bottleneck,'' \emph{arXiv preprint arXiv:1612.00410}, 2016.

\bibitem{kingma2022autoencoding}
D.~P. Kingma and M.~Welling, ``Auto-encoding variational bayes,'' 2022.

\bibitem{Tishby2000}
N.~Tishby, F.~C. Pereira, and W.~Bialek, ``The information bottleneck method,'' 2000.

\bibitem{boyd2004convex}
S.~P. Boyd and L.~Vandenberghe, \emph{Convex optimization}.\hskip 1em plus 0.5em minus 0.4em\relax Cambridge university press, 2004.

\bibitem{9145068}
J.~Shao and J.~Zhang, ``Bottlenet++: An end-to-end approach for feature compression in device-edge co-inference systems,'' in \emph{2020 IEEE International Conference on Communications Workshops (ICC Workshops)}, 2020, pp. 1--6.

\bibitem{Gaussian_ep_prior1}
H.~Li, J.~Shao, H.~He, S.~Song, J.~Zhang, and K.~B. Letaief, ``Tackling distribution shifts in task-oriented communication with information bottleneck,'' \emph{IEEE Journal on Selected Areas in Communications}, vol.~43, no.~7, pp. 2667--2683, 2025.

\bibitem{Gaussian_ep_prior2}
F.~Binucci, P.~Banelli, P.~Di~Lorenzo, and S.~Barbarossa, ``Opportunistic information-bottleneck for goal-oriented feature extraction and communication,'' \emph{IEEE Open Journal of the Communications Society}, vol.~5, pp. 2418--2432, 2024.

\bibitem{ep_priorloguni}
S.~Xie, S.~Ma, M.~Ding, Y.~Shi, M.~Tang, and Y.~Wu, ``Robust information bottleneck for task-oriented communication with digital modulation,'' \emph{IEEE Journal on Selected Areas in Communications}, vol.~41, no.~8, pp. 2577--2591, 2023.

\bibitem{takahashi2019variational}
H.~Takahashi, T.~Iwata, Y.~Yamanaka, M.~Yamada, and S.~Yagi, ``Variational autoencoder with implicit optimal priors,'' in \emph{Proceedings of the AAAI Conference on Artificial Intelligence}, vol.~33, 2019, pp. 5066--5073.

\bibitem{Sugiyama_Suzuki_Kanamori_2012}
M.~Sugiyama, T.~Suzuki, and T.~Kanamori, \emph{Density Ratio Estimation in Machine Learning}.\hskip 1em plus 0.5em minus 0.4em\relax Cambridge University Press, 2012.

\bibitem{bertsekas1997nonlinear}
D.~P. Bertsekas, ``Nonlinear programming,'' \emph{Journal of the Operational Research Society}, vol.~48, no.~3, pp. 334--334, 1997.

\bibitem{kingma2015variational}
D.~P. Kingma, T.~Salimans, and M.~Welling, ``Variational dropout and the local reparameterization trick,'' \emph{Advances in neural information processing systems}, vol.~28, 2015.

\bibitem{molchanov2017variational}
D.~Molchanov, A.~Ashukha, and D.~Vetrov, ``Variational dropout sparsifies deep neural networks,'' in \emph{International conference on machine learning}.\hskip 1em plus 0.5em minus 0.4em\relax PMLR, 2017, pp. 2498--2507.

\bibitem{ImageRecoveryClassification}
Z.~Lyu, G.~Zhu, J.~Xu, B.~Ai, and S.~Cui, ``Semantic communications for image recovery and classification via deep joint source and channel coding,'' \emph{IEEE Transactions on Wireless Communications}, vol.~23, no.~8, pp. 8388--8404, 2024.

\bibitem{LatentDiffusion}
J.~Pei, C.~Feng, P.~Wang, H.~Tabassum, and D.~Shi, ``Latent diffusion model-enabled low-latency semantic communication in the presence of semantic ambiguities and wireless channel noises,'' \emph{IEEE Transactions on Wireless Communications}, vol.~24, no.~5, pp. 4055--4072, 2025.

\bibitem{ExplainableSemantic}
S.~Ma, W.~Qiao, Y.~Wu, H.~Li, G.~Shi, D.~Gao, Y.~Shi, S.~Li, and N.~Al-Dhahir, ``Task-oriented explainable semantic communications,'' \emph{IEEE Transactions on Wireless Communications}, vol.~22, no.~12, pp. 9248--9262, 2023.

\bibitem{RobustInformationBottleneck}
S.~Xie, S.~Ma, M.~Ding, Y.~Shi, M.~Tang, and Y.~Wu, ``Robust information bottleneck for task-oriented communication with digital modulation,'' \emph{IEEE Journal on Selected Areas in Communications}, vol.~41, no.~8, pp. 2577--2591, 2023.

\bibitem{Multi-DeviceTask-Oriented}
C.~Cai, X.~Yuan, and Y.-J. Angela~Zhang, ``Multi-device task-oriented communication via maximal coding rate reduction,'' \emph{IEEE Transactions on Wireless Communications}, vol.~23, no.~12, pp. 18\,096--18\,110, 2024.

\bibitem{deng2012mnist}
L.~Deng, ``The mnist database of handwritten digit images for machine learning research,'' \emph{IEEE Signal Processing Magazine}, vol.~29, no.~6, pp. 141--142, 2012.

\bibitem{cifar}
A.~Krizhevsky, ``Learning multiple layers of features from tiny images,'' Tech. Rep., 2009.

\bibitem{cover1999elements}
T.~M. Cover, \emph{Elements of information theory}.\hskip 1em plus 0.5em minus 0.4em\relax John Wiley \& Sons, 1999.

\bibitem{pinsker1964}
M.~S. Pinsker, ``Information and information stability of random variables and processes,'' \emph{Holden-Day}, 1964.

\bibitem{vaeapproxerror}
A.~Shekhovtsov, D.~Schlesinger, and B.~Flach, ``Vae approximation error: Elbo and exponential families,'' \emph{arXiv preprint arXiv:2102.09310}, 2021.

\bibitem{kingma2019introduction}
D.~P. Kingma, M.~Welling \emph{et~al.}, ``An introduction to variational autoencoders,'' \emph{Foundations and Trends{\textregistered} in Machine Learning}, vol.~12, no.~4, pp. 307--392, 2019.

\bibitem{dai2019diagnosing}
B.~Dai and D.~Wipf, ``Diagnosing and enhancing vae models,'' \emph{arXiv preprint arXiv:1903.05789}, 2019.

\end{thebibliography}

\begin{IEEEbiography}[{\includegraphics[width=1in,height=1.25in,clip,keepaspectratio]{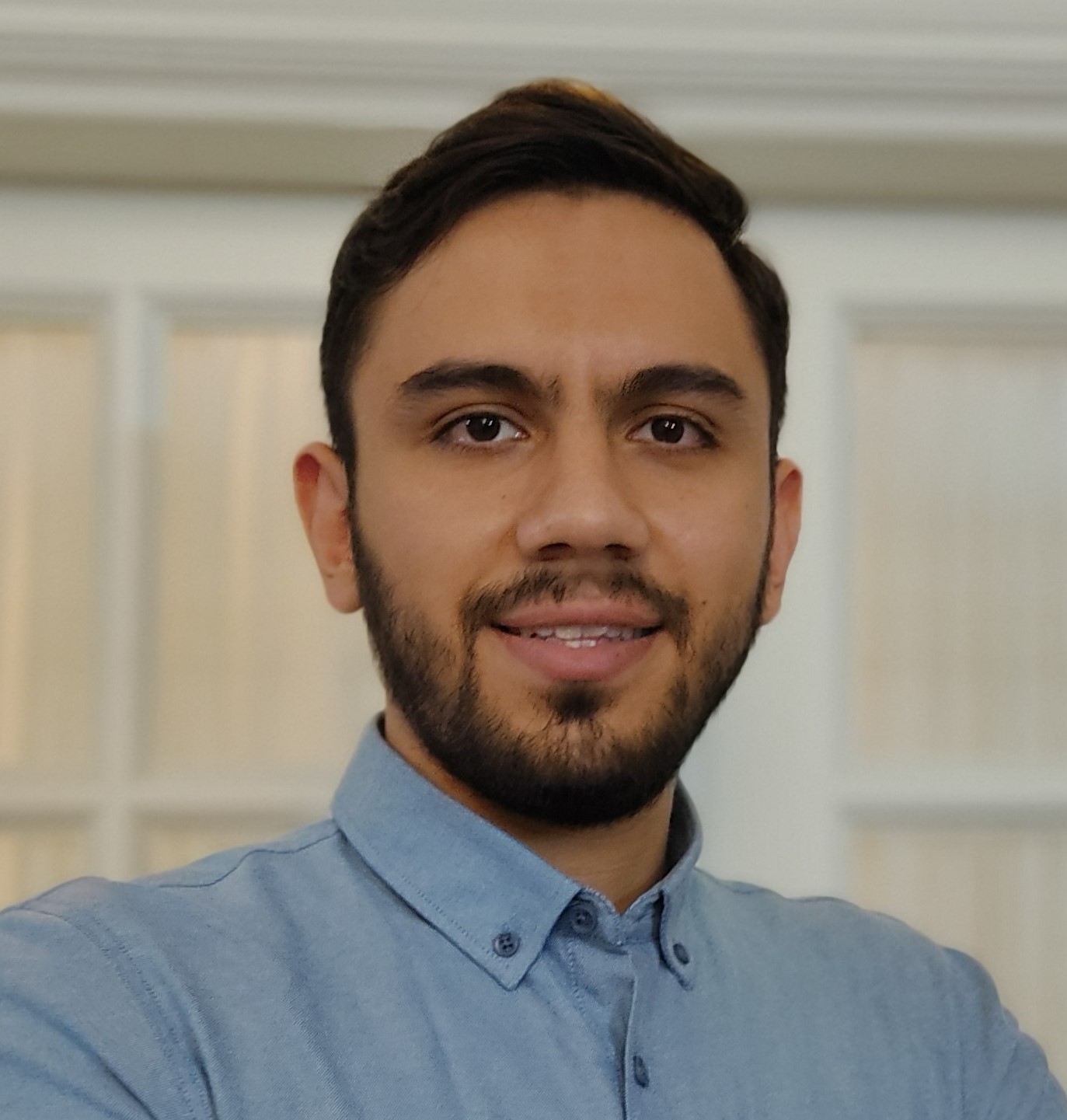}}]{Ahmad Halimi Razlighi } (Graduate Student Member, IEEE) received his B.Sc. in Electrical Engineering-Telecommunication in 2018 as the top graduate student from the Iran Broadcasting University (IRIBU), Tehran, Iran. In 2022, he graduated from the Iran University of Science and Technology (IUST), Tehran, Iran, with an M.Sc. in Electrical Engineering-Communication-Systems. He is currently with the Department of Communications Engineering (ANT) at the University of Bremen, Germany, pursuing his Ph.D. in Communication Engineering. His research interests include semantic communication, task-oriented communication, and machine learning for wireless communication.
\end{IEEEbiography}

\begin{IEEEbiography}[{\includegraphics[width=1in,height=1.25in,clip,keepaspectratio]{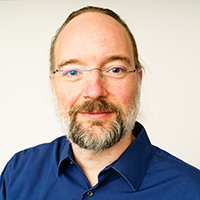}}]{Carsten Bockelmann } (Member, IEEE) received his Dipl.-Ing and Ph.D. degrees in Electrical Engineering from the University of Bremen, Germany in 2006 and 2012 respectively. Since 2012 he has been with the Department of Communications Engineering (ANT) as a Senior Research Group Leader and his main research interests include but are not limited to applications of compressive sensing and machine learning in communication, massive machine-type communication, ultra-reliable low latency communication, compressive sampling, and channel coding.
\end{IEEEbiography}

\begin{IEEEbiography}[{\includegraphics[width=1in,height=1.25in,clip,keepaspectratio]{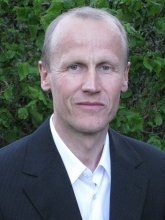}}]{Armin Dekorsy }
(Senior Member, IEEE) is a Professor with the University of Bremen, where he is the Director of the Gauss-Olbers Space Technology Transfer Center and Heads the Department of Communications Engineering. With over 11 years of industry experience, including distinguished research positions, such as a DMTS with Bell Labs and a Research Coordinator Europe with Qualcomm, he has actively participated in more than 65 international research projects, with leadership roles in 17 of them. He co-authored the textbook Nachrichtenübertragung (Release 6, Springer Vieweg), which is a bestseller in the field of communication technologies in Germanspeaking countries. His research focuses on signal processing and wireless communications for 5G/6G, industrial radio, and 3-D networks. He is a Senior Member of the IEEE Communications and Signal Processing Society and a member of the VDE/ITG Expert Committee on Information and System Theory.
\end{IEEEbiography}

\end{document}